\pdfoutput=1 
\documentclass[aps,prl,twocolumn,superscriptaddress,reprint,floatfix]{revtex4-2}  %
\usepackage{graphicx}  %
\usepackage{caption}
\usepackage{dcolumn}   %
\usepackage{bm}        %
\usepackage{bbm}       %
\usepackage{amssymb}   %
\usepackage{amsmath}   %
\usepackage{siunitx} 
\usepackage{color}
\usepackage{here}
\usepackage{booktabs}
\usepackage{multirow}
\usepackage{subcaption}
\usepackage{placeins}

\usepackage{xr} %
\makeatletter
\newcommand*{\addFileDependency}[1]{%
  \typeout{(#1)}
  \@addtofilelist{#1}
  \IfFileExists{#1}{}{\typeout{No file #1.}}
}
\makeatother

\newcommand*{\myexternaldocument}[1]{%
    \externaldocument[S-]{#1}%
    \addFileDependency{#1.tex}%
    \addFileDependency{#1.aux}%
}
\myexternaldocument{supplementary2}

\makeatletter
\@ifclasswith{revtex4-1}{linenumbers}{\newcommand{\equationlinenobegin}{\begin{linenomath}}}{\newcommand{\equationlinenobegin}{}}
\@ifclasswith{revtex4-1}{linenumbers}{\newcommand{\equationlinenoend}{\end{linenomath}}}{\newcommand{\equationlinenoend}{}}
\makeatother

\usepackage[breaklinks=true,pdfborder={0 0 0},pdftex]{hyperref}%

\newcommand{\ave}[1]{\left< {#1} \right>}

\newcommand{\cum}[2]{\left< {#1}^{#2} \right>_c}
\newcommand{\Sk}[1]{\mathrm{Sk}\left[{#1}\right]}
\newcommand{\Ku}[1]{\mathrm{Ku}\left[{#1}\right]}
\newcommand{\Cov}[2]{\mathrm{Cov}[#1,#2]}

\newcommand{\betakpz}{{\beta^\mathrm{KPZ}}}
\newcommand{\zkpz}{{z^\mathrm{KPZ}}}
\newcommand{\xikpz}[1]{{\xi^\mathrm{KPZ}({#1})}}

\newcommand{\betadp}{{\beta^\mathrm{DP}}}

\newcommand{\betacdp}{{\beta^\mathrm{CDP}}}
\newcommand{\betaprimecdp}{{{\beta'}^\mathrm{CDP}}}

\newcommand{\nupara}{{\nu_{\parallel}}}
\newcommand{\nuperp}{{\nu_{\perp}}}

\newcommand{\zcdp}{{z^\mathrm{CDP}}}
\newcommand{\zabs}{{z^\mathrm{(C)DP}}}

\newcommand{\nuparadp}{{\nu^\mathrm{DP}_{\parallel}}}
\newcommand{\nuperpdp}{{\nu^\mathrm{DP}_{\perp}}}
\newcommand{\nuparacdp}{{\nu^\mathrm{CDP}_{\parallel}}}
\newcommand{\nuperpcdp}{{\nu^\mathrm{CDP}_{\perp}}}

\newcommand{\rhoval}{{\rho(\mathbf{x},t)}}

\newcommand{\rhoinf}{{\rho_\infty(\epsilon)}}

\newcommand{\spanwiserho}{{\left<\rhoval\right>_x}}
\newcommand{\xipara}{{\xi_{\parallel}(\epsilon)}}
\newcommand{\xiperp}{{\xi_{\perp}(\epsilon)}}
\newcommand{\xidyn}[1]{{\xi_{\mathrm{dyn}} \left({#1}\right) }}

\begin{document}

\title{Universal interface fluctuations in absorbing-state phase transitions}

\author{Yohsuke T. Fukai}
\email{ysk@yfukai.net}
\affiliation{%
  Nonequilibrium Physics of Living Matter Laboratory, 
  RIKEN Pioneering Research Institute,  
  2-2-3 Minatojima-minamimachi, Chuo-ku, Kobe, Hyogo 650-0047, Japan
}
\affiliation{%
  Nonequilibrium Physics of Living Matter Riken Hakubi Research Team, 
  RIKEN Center for Biosystems Dynamics Research, 
  2-2-3 Minatojima-minamimachi, Chuo-ku, Kobe, Hyogo 650-0047, Japan
}%
\affiliation{%
	Department of Physics, The University of Tokyo, 7-3-1 Hongo, Bunkyo-ku, Tokyo 113-0033, Japan
}%
\author{Keiichi Tamai}%
\affiliation{%
    Department of Physics, The University of Tokyo, 7-3-1 Hongo, Bunkyo-ku, Tokyo 113-0033, Japan
}%
\affiliation{%
  Institute for Physics of Intelligence, The University of Tokyo, 7-3-1 Hongo, Bunkyo-ku, Tokyo 113-0033, Japan
}%
\affiliation{%
  The Institute for Solid State Physics, The University of Tokyo, 5-1-5 Kashiwanoha, Kashiwa, Chiba 277-8581, Japan
}%
\author{Tetsuya Hiraiwa}%
\affiliation{%
  Department of Physics, The University of Tokyo, 7-3-1 Hongo, Bunkyo-ku, Tokyo 113-0033, Japan
}%
\affiliation{%
  Institute of Physics, Academia Sinica, No. 128, Section 2, Academia Road, Taipei 11529, Taiwan
}%
\affiliation{%
  Physics Division, National Center for Theoretical Sciences, Taipei 106319, Taiwan
}%

\affiliation{%
  Mechanobiology Institute, National University of Singapore, 5A Engineering Drive 1, Singapore 117411, Singapore
}%
\affiliation{%
  Universal Biology Institute, The University of Tokyo, 7-3-1 Hongo, Bunkyo-ku, Tokyo 113-0033, Japan
}%

\date{\today}

\begin{abstract}
Despite similarities between models exhibiting absorbing phase transitions (APTs) and those showing Kardar-Parisi-Zhang (KPZ) growth, the relationship between these universal fluctuations has remained elusive.  We numerically study (1+1)-dimensional interfaces of (2+1)-dimensional models showing APTs of directed percolation (DP) and compact directed percolation (CDP) classes with an active boundary, finding a universal crossover from short-time APT-governed fluctuations to long-time KPZ fluctuations.
Upon rescaling time and length by the APT correlation time and length, the cumulants of the interface height distributions collapse onto a single scaling function. The fluctuation properties of the discrete Domany-Kinzel model and the continuum stochastic Fisher-Kolmogorov-Petrovsky-Piskunov (sFKPP) equation coincide, indicating that the KPZ growth parameters are determined solely by fundamental properties of the APT. 
For the CDP sFKPP equation, a dimensionless parameter tunes both the critical interface distribution and the KPZ parameters, with the interface properties of the biased voter model recovered in a limiting case.
These results uncover a universal crossover in which KPZ fluctuations emerge from APT fluctuations at long times, linking paradigmatic universality classes of nonequilibrium scale-invariant phenomena.
\end{abstract}

\maketitle

\paragraph{Introduction.}

Emergent scale invariance often leads to universal large-scale properties of fluctuations in nonequilibrium systems. Absorbing phase transitions (APTs) and interface growth are representative examples of such universality \cite{hinrichsen_2000,henkel_2009,takeuchi_2014,barabasi_stanley_1995,corwin_2012,takeuchi_2018}, and appear in various phenomena ranging from intermittent turbulence \cite{avila_2011,lemoult_2016,sano_2016} to population dynamics \cite{harris_1974,eden_1961,williams_1972,wakita_1997_selfaffinity}, which can be described by local growth and competition. 
APTs, nonequilibrium phase transitions into absorbing states from which no transition to another state is possible \cite{hinrichsen_2000,henkel_2009,takeuchi_2014}, are examples where \textit{bulk} scale invariance at the transition point leads to universal scaling laws for fluctuations. Here, the order parameter $\rho(\mathbf{x},t)$ evolves in $d$-dimensional space $\mathbf{x}\in\mathbb{R}^d$, and the system including the time axis is referred to as $(d+1)$-dimensional. The directed percolation (DP) class and the compact directed percolation (CDP) class are known as prototypical classes for cases with a single absorbing state and two symmetric absorbing states, respectively \cite{hinrichsen_2000,henkel_2009}. On the other hand, when active clusters propagate and form fluctuating $\left[\left(d-1\right)+1\right]$-dimensional \textit{interfaces} \cite{barabasi_stanley_1995,edwards_1982}, whose height $h(\mathbf{x},t)$, i.e., the coarse-grained position of such an interface at a dimensionally reduced spatial coordinate $\mathbf{x}\in\mathbb{R}^{d-1}$, often exhibits scale-invariant universal fluctuations which fall in a universality class. The Kardar-Parisi-Zhang (KPZ) universality class \cite{kardar_1986,barabasi_stanley_1995,corwin_2012,takeuchi_2014,takeuchi_2018} describes cases without special symmetries and conservation laws, and is found to be relevant in various systems \cite{takeuchi_2014}. Although each class is by now extensively studied, whether and how the two universalities are related has remained largely unaddressed.

A useful starting point is that several models are known to host both universal fluctuations in different parameter regimes.
One of the earliest observations dates back at least to 1972, when Williams and Bjerknes numerically studied \cite{williams_1972} 
a version of the biased voter model, a model known to show the CDP-class transition, and discussed its connection to the Eden model in the off-critical limit \cite{eden_1961}, which describes interface growth in the KPZ class. 
Similarly, the off-critical limit of the contact process \cite{harris_1974,hinrichsen_2000,henkel_2009}, a model showing the transition of the DP class, is identical to the Eden model \cite{kaya_2000}. 
Another example is the stochastic Fisher-Kolmogorov-Petrovsky-Piskunov (sFKPP) equation \cite{mueller_1994a,mueller_1995,tribe_1996, Doering_2003,pechenik_1999,nesic_2014} [Eq.~\eqref{eq:DP-CDP-Langevin}], which shows the DP- and CDP-class APTs \cite{mueller_1994a,henkel_2009}. 
Interface fluctuations of discrete and continuous models associated with the $(2+1)$-dimensional sFKPP equation have been studied, and it has been established \cite{kaya_2000,moro_2001,nesic_2014} that when the parameter is far from the critical value, the propagating wave front shows the fluctuations of the $(1+1)$-dimensional KPZ universality class. 

In each of these examples, the two universalities live in distinct parameter regimes of a single model: APT criticality at the transition point, and KPZ interface growth far from it. When the control parameter is close to the critical point, one may expect an interface growth process characterized by both APT and KPZ universality \cite{kaya_2000}.
However, no clear connection has been established between the universal properties of APT bulk fluctuations and those of the KPZ interface, nor between their characteristic temporal and spatial scales.

In this Letter, by simulating discrete and continuum models showing the DP- and CDP-class transitions, we reveal that the universality of APTs imposes strong constraints on the interface fluctuations of the noisy traveling waves. 
By rescaling length and time by the characteristic length and time of APTs, we find a universal crossover from a regime characterized by the APT exponents to that associated with the KPZ scaling, interconnecting the universal fluctuations.

\begin{figure*}[tb!]
 \centering
 \includegraphics[width=7in]{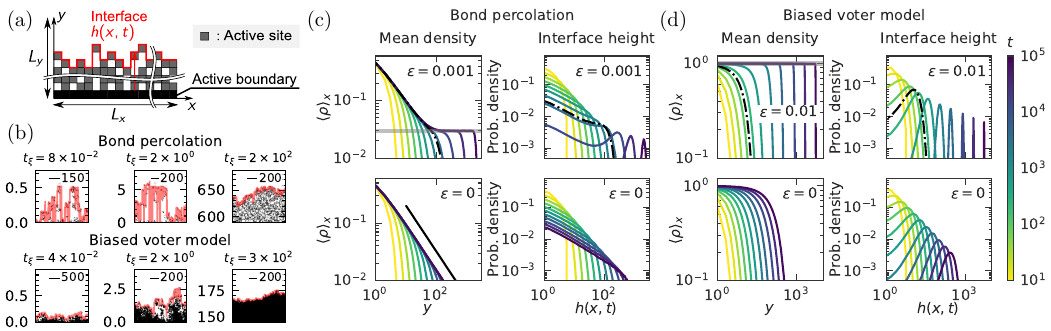}
 \caption{ Schematic and qualitative illustration of the interface growth process in the discrete models. (a) An illustration of the height definition for the discrete models with the active boundary.
 (b) Example snapshots of the active sites (black field) and the interface (red line) for the bond percolation (upper) and biased voter model (lower), respectively. 
 The axis label and the scale bar indicate the length rescaled by the correlation length $\xiperp$ and the actual length, respectively. The rescaled time $t_\xi=t/\xipara$ is denoted in each plot. The values of $\epsilon$ are $1\times10^{-5}$, $1\times10^{-4}$ and $5\times10^{-3}$ for the bond percolation and $1\times10^{-6}$, $5\times10^{-5}$ and $5\times10^{-3}$ for the biased voter model, respectively. 
 (c,d) The spanwise-averaged density $\spanwiserho$ (left) and height probability distribution (right) for the bond percolation (c) and biased voter model (d), respectively. 
 The top and bottom plots are the off-critical and critical cases, respectively. %
 The dash-dot lines are the data at $t\approx\xipara$.
 The gray lines in the off-critical mean-density plots indicate $\rhoinf$ (bond percolation) and $1$ (biased voter model). 
 The black line in the left bottom plot of (c) indicates the power law $\spanwiserho\propto{}y^{-\betadp/\nuperp}$.}
\label{fig:intro}
\end{figure*}

\paragraph{Models.}
To investigate model-independent properties of interface fluctuations associated with APTs, we studied discrete and continuum models showing the $(2+1)$-dimensional DP- and CDP-class transitions.
As the discrete models, we employed the bond percolation \cite{hinrichsen_2000,henkel_2009,wang_2013} for the DP class and the biased voter model (also called the Williams-Bjerknes model \cite{williams_1972}) for the CDP class, 
which can be defined as a special case of the Domany-Kinzel (DK) model \cite{domany_1984,hinrichsen_2000,henkel_2009} 
on the (2+1)-dimensional body-centered cubic (BCC) lattice \cite{wang_2013} (Fig.~\ref{S-fig:DK_schematic} \cite{supplementary}).
In the model, the sites $\rho((i,j),t)$ on a square lattice $i=0,\dots,L_x-1$, $j=0,\dots,L_y-1$ at the time $t$
take either the active [$\rho((i,j),t)=1$] or inactive [$\rho((i,j),t)=0$] state. The transition probability is determined by the percolation probability $p$ and the bias $\epsilon$ in the bond percolation and the biased voter model, respectively, as described in the End Matter \cite{supplementary}.

As the continuum models, we simulated the sFKPP equation for the (2+1)-dimensional DP and CDP classes, 
\equationlinenobegin\begin{equation}
\label{eq:DP-CDP-Langevin}
\frac{\partial \rho(\mathbf{x}, t)}{\partial t}=D \triangle\rho+A\rho-B\rho^2+\sigma\sqrt{f(\rho)}\eta(\mathbf{x}, t) \ ,
\end{equation}\equationlinenoend
where $B=1$ ($A=B=\epsilon$) and $f(\rho)=\rho$ [$f(\rho)=\rho(1-\rho)$] for the DP (CDP, respectively) class.
We used the algorithm in Ref.~\cite{dornic_2005} which ensures the nonnegativity of the density field. In this study, we used the lattice size $\Delta x=3$ and timestep size $\Delta t=0.25$, with the system size written as $L_x\Delta x\times L_y\Delta x$ \cite{supplementary}. 

Each model undergoes an absorbing-state phase transition at a critical parameter value [$p=p_c$ (bond percolation), $A=A_c$ (DP Langevin), and $\epsilon=0$ (CDP models); see Table \ref{tab:critical_parameters} and the End Matter for details \cite{supplementary}].
For the bond percolation and DP Langevin equation, we defined the distance from the transition point as $\epsilon:=\left(p-p_c\right)/p_c$ and $\epsilon:=\left(A-A_c\right)/A_c$, respectively. 
Through simulations from homogeneous initial conditions, we further estimated the APT correlation time $\xipara\sim|\epsilon|^{-\nupara}$ and correlation length
$\xiperp\sim|\epsilon|^{-\nuperp}$, which diverge as $\epsilon\to 0$  (Table~\ref{tab:exponent_values}; See End Matter and Figs.~\ref{S-fig:DP_mean_rho_A_c}-\ref{S-fig:CDP_corr_len_summarized} for the detailed methods and Tables~\ref{S-tab:DP_nonuniversal} and \ref{S-tab:CDP_nonuniversal} for the estimated parameters \cite{supplementary}).

\paragraph{Interface fluctuation with active boundary.}
We simulated the models with active walls to study the interface fluctuation [Figs.~\ref{fig:intro}(a) and \ref{S-fig:intro-langevin}(a) \cite{supplementary}]. Specifically, the sites at the wall ($y=0$) were always set to active ($\rhoval=\rho_b>0$ for the Langevin equations, where $\rho_b$ is a constant), and all other sites were initially set to inactive, after which the active region invaded [Figs.~\ref{fig:intro} and \ref{S-fig:intro-langevin} \cite{supplementary}].
We defined the interface as the distance between the wall and the furthest active site (a lattice site with $\rhoval>\rho_{th}$ for the Langevin equations \cite{mueller_1995,tribe_1996}) [Figs.~\ref{fig:intro}(a,b) and \ref{S-fig:intro-langevin}(a,b) \cite{supplementary}]. 

The spanwise-averaged density showed a crossover from the critical-regime profile to that of a traveling wave around $t\sim\xipara$ [Figs.~\ref{fig:intro}(c) and \ref{S-fig:intro-langevin}(c) \cite{supplementary}], confirming that $\xipara$ is also the characteristic timescale for the interface dynamics; the interface morphology and height distribution function likewise qualitatively changed around $\xipara$ [Figs.~\ref{fig:intro}(b,d) and \ref{S-fig:intro-langevin}(b,d) \cite{supplementary}].

To quantify this crossover, we calculated the cumulants $\cum{h(x,t)}{k}$ ($k=1,2,3,4$) for the discrete models and found consistent behavior across $\epsilon$ values (Fig.~\ref{S-fig:raw_cumulants_DK} \cite{supplementary}). Interfaces at criticality ($\epsilon=0$) exhibited power-law growth with the APT dynamic exponent $\cum{h}{k} \sim t^{k/\zabs}$, while for large $\epsilon$ a temporal region emerged in which the KPZ-class exponents were observed; $\cum{h}{k}\sim t\;(k=1)$ and $\cum{h}{k}\sim t^{k\betakpz}\;(k>1)$, where $\betakpz=1/3$ is the growth exponent for the KPZ class (Table~\ref{tab:exponent_values} \cite{supplementary}). This KPZ regime appeared at progressively earlier times as $\epsilon$ increased.
These observations motivate rescaling the height, time, and spatial coordinate by the characteristic length and temporal scales of the APT, $\xiperp$, $\xipara$, and $\xiperp$, respectively.

\paragraph{Universal crossover from APT to KPZ-class fluctuation.}

\begin{figure}[tb!]
 \centering
 \includegraphics[width=3.2in]{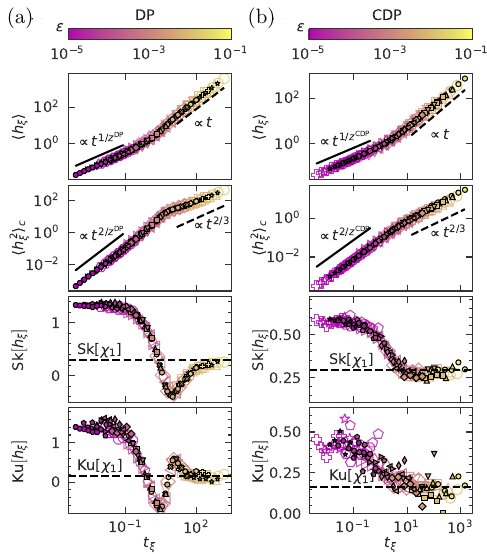}
 \caption{\label{fig:rescaled_cumulants} Cumulants and cumulant ratios of the height distribution function for the (a) DP- and (b) CDP-class models. The plots show the mean, variance, skewness, and kurtosis for the rescaled height ($\ave{h_\xi}$, $\cum{h_\xi}{2}$, $\Sk{h_\xi}$ and $\Ku{h_\xi}$) for the off-critical cases. The larger unfilled and smaller filled symbols are for the discrete models and the Langevin equations [$D=1$, $\sigma=1$, $\rho_b=0.1$, and $\rho_{th} = 0.1$ (DP) and $D=0.1$, $\sigma=1$, $\rho_b=1$, and $\rho_{th}=0$ (CDP)], respectively. 
 The solid and dashed lines in the upper two plots are guides for the eye for the APT and KPZ exponents (see the main text), respectively. The dashed lines in the lower two plots show the theoretical values for the GOE-TW distribution, $\chi_1$.}
\end{figure}

We rescaled the height, time, and spatial coordinate as
\begin{equation}
h_\xi(x_\xi,t_\xi):=h(x,t)/\xiperp,\;t_\xi:=t/\xipara,\;x_\xi:=x/\xiperp,
\end{equation}
and found that the cumulants collapse onto unique scaling functions bridging the APT and KPZ regimes (Fig.~\ref{fig:rescaled_cumulants}).
These functions show the asymptotic APT scaling laws in the region $t_\xi\ll 1$ (the short-time regime) and the KPZ scaling laws in $t_\xi\gg 1$ (the long-time regime), suggesting 
\begin{equation}
\cum{h_\xi(x_\xi,t_\xi)}{k} \simeq g_k(t_\xi),\; g_k(t_\xi)\sim \begin{cases}
t_\xi^{k/\zabs} (t_\xi\ll 1)\\
t_\xi \;\;\;\;\;\;(t_\xi\gg 1, k=1) \\
t_\xi^{k\betakpz} (t_\xi\gg 1, k>1). 
\end{cases}
\end{equation}
Interfaces at criticality ($\epsilon=0$) exhibited the same scaling laws as those observed in the short-time limit (Fig.~\ref{S-fig:raw_cumulants_DK} \cite{supplementary}). 

The skewness $\Sk{h_\xi}:=\cum{h_\xi}{3}/\cum{h_\xi}{2}^{3/2}$ and the kurtosis $\Ku{h_\xi}:=\cum{h_\xi}{4}/\cum{h_\xi}{2}^{2}$ exhibited plateaus in both the short- and long-time limits (Fig.~\ref{fig:rescaled_cumulants}), indicating the existence of asymptotic height distributions in each regime.
In the short-time regime, the plateau values agreed with those obtained at criticality ($\epsilon=0$), where both quantities remained constant over a long timescale (Fig.~\ref{S-fig:critical_skew_kurtosis} \cite{supplementary}), suggesting that the height distribution converges to a well-defined critical distribution.
In the long-time regime, both values converged to those of the Gaussian orthogonal ensemble Tracy--Widom (GOE-TW) distribution, the exact solution for the KPZ class with a flat initial condition. This convergence further confirmed the emergence of the KPZ-class fluctuation. 

\begin{figure}[tb!]
 \centering
 \includegraphics[width=3.2in]{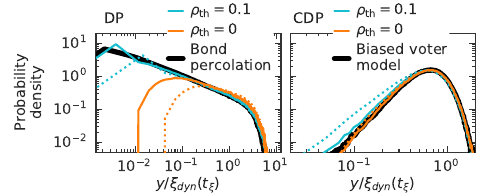}
 \caption{
  The critical height distribution for the (left) DP-class and (right) CDP-class models.
  Data for the discrete models (black solid lines) and the Langevin equations [$\rho_b=0.1$ for the left and $\rho_b=1$ for the right] are shown. The dashed and solid curves show the data at $t=1\times10^4$ and $t=1\times10^5$, respectively.}
 \label{fig:histgrams}
\end{figure}

\paragraph{Universality of height fluctuation.}
Comparison between the discrete models and the Langevin equations revealed the universality of the height fluctuation across models.
At criticality ($\epsilon=0$), the height distributions of the Langevin equations, rescaled by the dynamic correlation length, asymptotically converged to those of the critical bond percolation, irrespective of the threshold $\rho_{th}$; the cutoffs at $y>0$ observed with smaller $\rho_{th}$ or larger $\rho_b$ values decreased as time increased [Figs.~\ref{fig:histgrams} (left) and \ref{S-fig:critical_height_distributions_DP} \cite{supplementary}], suggesting the existence of a universal asymptotic height distribution for models in the DP class.
Similarly, for the CDP Langevin equation, the moderate effect of $\rho_{th}$ disappeared in the long-time limit [Figs.~\ref{fig:histgrams} (right) and \ref{S-fig:critical_height_distributions_CDP} \cite{supplementary}].
At $D/\sigma^2 \ll 1$, the distribution function overlapped with that for the biased voter model, confirming the universality of the critical height distribution (Figs.~\ref{fig:histgrams_CDP} and \ref{S-fig:critical_height_distributions_CDP_shift_scale} \cite{supplementary}). Interestingly, at larger $D/\sigma^2$, the asymptotic height distribution retained almost the same shape but was shifted (Figs.~\ref{fig:histgrams_CDP}, \ref{S-fig:critical_height_distributions_CDP} and \ref{S-fig:critical_height_distributions_CDP_shift_scale} \cite{supplementary}), suggesting that the dimensionless parameter $D/\sigma^2$, which is marginal at $d=2$ under rescaling of space and time, has a fundamental effect on the critical height distribution even after rescaling by the diverging CDP correlation length.

This universality extended to the off-critical regime. For $\epsilon>0$, the cumulants of the Langevin equations collapsed onto the same scaling functions as those of the discrete models under conditions where the critical height distributions matched (Figs.~\ref{fig:rescaled_cumulants} and \ref{S-fig:raw_cumulants_langevin} \cite{supplementary}). For the CDP Langevin equations, the cumulant scaling functions depended slightly on $D/\sigma^2$, further supporting the role of $D/\sigma^2$ as a fundamental parameter governing the large-scale behavior (Fig.~\ref{S-fig:CDP_cumulants_overlap} \cite{supplementary}).

\paragraph{KPZ growth with universal parameters.}

In the KPZ regime, the asymptotic height is described by 
\begin{equation} \label{eq:h_asymptotic}
h_\xi(x_\xi,t_\xi) \simeq  v_\infty t_\xi + (\Gamma t_\xi)^{\betakpz} \chi(X)\;\;(t_\xi \rightarrow \infty),
\end{equation}
with $X:=x_\xi/\xikpz{t_\xi}$ and $\xikpz{t_\xi}:=\frac{2}{A}(\Gamma{}t_\xi)^{1/\zkpz}$, where $\zkpz=3/2$ is the dynamical exponent for the KPZ class (Table~\ref{tab:exponent_values} \cite{supplementary}), $\chi(X)$ is the universal stochastic variable, and $v_\infty$, $\Gamma$, and $A$ are system-dependent parameters \cite{takeuchi_2018,prahofer_2000,takeuchi_2010}. These parameters can be estimated from the height cumulants following standard methods \cite{takeuchi_2012,takeuchi_2018,krug_1992_amplitude} (See the End Matter \cite{supplementary}).

Owing to the universality of the height fluctuation scaled by the APT correlation time and length, we expect these parameter values to be universal, in the sense that they only depend on the fundamental properties of the APT irrespective of the system details.
We found that the estimated parameters for the bond percolation and DP Langevin equation indeed agree within the uncertainties, supporting the idea of KPZ growth with universal parameters (Fig.~\ref{S-fig:KPZ_nonuniversal} and Tables \ref{tab:nonuniversal_params_KPZ_full} and \ref{S-tab:nonuniversal_params_KPZ} \cite{supplementary}). For the CDP models, in accordance with the observation that the critical height distribution depends on the dimensionless parameter $D/\sigma^2$, the KPZ parameters were also dependent on $D/\sigma^2$ for the Langevin equation and agreed with those for the biased voter model with small $D/\sigma^2$, suggesting its universality (Fig.~\ref{S-fig:KPZ_nonuniversal} and Table \ref{S-tab:nonuniversal_params_KPZ} \cite{supplementary}). 
In the original scale of $h(x,t)$, the parameters, denoted by $v_\infty'$, $\Gamma'$, and $A'$, follow
\begin{equation}
	v'_\infty = v_\infty \frac{\xipara}{\xiperp}, \; \;
	\Gamma' = \Gamma \frac{\xiperp^3}{\xipara},\;\; 
	A' = A \xiperp, 
\end{equation}
which leads to the power laws for the KPZ parameters
$v_\infty' \sim \epsilon^{\nu_\perp-\nu_\parallel},\; \Gamma' \sim \epsilon^{\nu_\parallel-3\nu_\perp} ,\; $ and $A' \sim \epsilon^{-\nu_\perp}$.

When rescaled by the dynamical correlation length $\xidyn{t}$ in the short-time regime and by the KPZ parameters as $\frac{h_\xi(x_\xi,t_\xi)-v_\infty t_\xi}{\left(\Gamma t_\xi\right)^{1/3}}$ in the long-time regime [Eq.~\eqref{eq:h_asymptotic}], the height distributions collapsed onto unique curves in both limits (Figs.~\ref{S-fig:off_critical_height_distributions_DP} and \ref{S-fig:off_critical_height_distributions_CDP} \cite{supplementary}). In the short-time regime, the height histograms converged to that of the critical interfaces ($\epsilon=0$). In the long-time regime, the height distribution agreed with the exact solution for the KPZ class with the flat initial condition, the GOE-TW distribution. We also found consistent behavior for the height spatial covariance $C_{s,int}(l,t_\xi):=\Cov{h_\xi(x_\xi+l,t_\xi)}{h_\xi(x_\xi,t_\xi)}$ (End Matter; Figs.~\ref{S-fig:off_critical_spatial_correlation_DP} and \ref{S-fig:off_critical_spatial_correlation_CDP} \cite{supplementary}).

\paragraph{Discussion and conclusion.}

We investigated the fluctuations of interfaces defined by the farthest active sites from the active boundary in the models exhibiting APTs of the DP and CDP classes. 
By rescaling the spatial and temporal scales by the APT correlation length and time, respectively, we found a universal crossover from height fluctuations characterized by the APT dynamical correlation length and the statistical properties of the critical interfaces to those of the KPZ class. The scaling function was found to be robust, with its long-time limit independent of the model details and of the threshold $\rho_{th}$ used to define the interface in the Langevin cases.
The result suggests that one can predict the statistical properties of the interfaces solely from the bulk correlation length and time, and conversely that properties of the critical fluctuation in the bulk can be inferred from the evolution of the interface fluctuation. Interestingly, for the CDP Langevin equation, the dimensionless parameter $D/\sigma^2$ tuned both the critical fluctuation properties and the KPZ-interface parameters. The results showed agreement with the biased voter model in the limit $D/\sigma^2 \to 0$.

Our results show that the KPZ fluctuation emerges from APT fluctuations in the long-time limit. It would be interesting to explore whether this crossover scenario of ``nested'' universal fluctuations can be generalized to other classes of APTs \cite{hinrichsen_2000,henkel_2009}. Furthermore, understanding the circumstances that determine the interface fluctuation would be important, given that the interface fluctuations associated with the Ginzburg-Landau equation have been reported to exhibit Edwards-Wilkinson-class fluctuations \cite{caballero_2020}. The effect of noise in the sFKPP equation has been extensively studied~\cite{nesic_2014}, and a unified understanding bridging our results with those obtained in its strong-noise limit~\cite{hallatschek_2009} would be another interesting direction. Another open question is whether the critical APT-associated interface fluctuation shows geometry dependence analogous to that of the KPZ class \cite{prahofer_2000,corwin_2012,takeuchi_2018,fukai_2020_kardarparisizhang}.

Experimental realizations in phenomena associated with APTs \cite{takeuchi_2007a,takeuchi_2014,sano_2016} or in neighbor interactions analogous to the voter model \cite{mesa_2018} would also be valuable to pursue. Notably, the DP-class APT and the KPZ-class interface growth have both been experimentally observed in similar setups of liquid crystal electroconvection \cite{takeuchi_2007a,takeuchi_2010,takeuchi_2014}, making this system a promising candidate to probe the crossover predicted here. The KPZ growth with universal parameters is expected at temporal and spatial scales $t$ and $l$ satisfying $t_\mathrm{micro}\ll\xipara\ll t$ and $l_\mathrm{micro}\ll\xiperp\ll l$, where $t_\mathrm{micro}$ and $l_\mathrm{micro}$ denote the microscopic time and length scales of the dynamics. This regime is accessible, for example, in the DSM1--DSM2 transition of electroconvecting liquid crystals \cite{takeuchi_2007a}, for which the APT correlation time and length, $\xipara \approx \SI{5}{\s}$ and $\xiperp \approx \SI{100}{\um}$ at $\epsilon=10^{-2}$, lie below the typical experimental observation windows \cite{takeuchi_2009}.

\begin{acknowledgments}
  We thank Hiroki Yamaguchi for the preliminary simulations and discussions on the exact expressions in the CDP cases.
  We acknowledge valuable discussion with Kazumasa A. Takeuchi, Kyogo Kawaguchi, Kyosuke Adachi, Rory Cerbus, Takeyuki Miyawaki and Somayeh Zeraati.
  We are grateful to Michael Pr{\"a}hofer and Herbert Spohn for the theoretical curves of the GOE Tracy--Widom distribution, which are made available online \cite{Prahofer.Spohn-Table}, and to Folkmar Bornemann for the curves of the Airy$_1$ covariance, evaluated by his algorithm in \cite{Bornemann-MC2010}.
  This work is supported in part by KAKENHI from Japan Society for the Promotion of Science (Grant Nos. JP17J05559, JP22K14016, 19H05795, and 25H01360) and JST ACT-X grant number JPMJAX24LF (to Y.T.F.) and the Seed fund of Mechanobiology Institute (to T.H.).
  We thank the Supercomputer Center of the Institute for Solid State Physics (the University of Tokyo), the Meiji Institute for Advanced Study of Mathematical Sciences (Meiji University) and the Center for Computational Sciences (University of Tsukuba) for computational resources.
\end{acknowledgments}

\appendix
\clearpage
\section*{End matter}
\renewcommand{\thefigure}{A\arabic{figure}}
\renewcommand{\thetable}{A\Roman{table}}
\renewcommand{\theequation}{A\arabic{equation}}

\setcounter{figure}{0}
\setcounter{table}{0}
\setcounter{equation}{0}

\begin{table}[!ht]
	\caption{\label{tab:exponent_values}
		Values of universal exponents \cite{hinrichsen_2000,henkel_2009}. For the DP class, we used the values estimated in Ref.~\cite{wang_2013}.}
        \begin{tabular}{l|c|c|c|c|c|c|c}
            \toprule
            Name & $\betadp$ & $\nuparadp$ & $\nuperpdp$ & $\betacdp$ & $\betaprimecdp$ & $\nuparacdp$ & $\nuperpcdp$\\
            \midrule
            Value & $0.580(4)$ & $1.287(2)$ & $0.729(1)$ & $0$ & $1$ & $1$ & $1/2$\\
		\bottomrule
	\end{tabular}
\end{table}

\begin{table}[!ht]
	\caption{\label{tab:critical_parameters}
		Values of the critical parameters for the DP models. For the bond percolation, we used the values estimated in \cite{wang_2013}.}
        \begin{tabular}{l|c|c}
            \toprule
            Model & Bond percolation & DP Langevin\\
            \midrule
			Name & $p_c$ & $A_c$ \\
            \midrule
			Value & $0.28733837(2)$ & $0.181358(4)$ \\
		\bottomrule
	\end{tabular}
\end{table}

\begin{table}[!ht]
	\caption{\label{tab:nonuniversal_params_KPZ_full}
		Estimated parameters of the KPZ class in the long-time regime (excerpt).
            For the Langevin equations, the data for $D=1,\sigma=1,\rho_b=0.1$ (DP) and $D=0.1,\sigma=1,\rho_b=1$ (CDP) are shown, respectively. 
		The ``Value'' column indicates the value estimated with the values of the APT nonuniversal parameters shown in Table~\ref{S-tab:DP_nonuniversal} and \ref{S-tab:CDP_nonuniversal}, 
		and the ``Max'' and ``Min'' columns indicate the maximum and minimum values obtained by varying the APT nonuniversal parameters within the uncertainties, respectively.
		In the ``Value'' column, the statistical uncertainties are shown in parentheses.
    The full table is shown in Table~\ref{S-tab:nonuniversal_params_KPZ} \cite{supplementary}. The values are also shown in Fig.~\ref{S-fig:KPZ_nonuniversal} \cite{supplementary} as a plot.
		}
		\begin{tabular}{c|c|c|c|c|c|c}
			\toprule
			   & \multicolumn{3}{c|}{$v_{\infty}$} & \multicolumn{3}{c}{$\Gamma$} \\
			\cline{2-7}
			model & Value & Max & Min & Value & Max & Min \\
			\midrule
			Bond percolation & $2.756(3)$ & $2.894$ & $2.627$ & $14.2(1)$ & $16.8$ & $12.1$ \\
			\midrule
			DP Langevin & $2.84(1)$ & $3.00$ & $2.70$ & $15.1(5)$ & $18.5$ & $12.5$ \\
			\midrule
			Biased voter & $0.539(3)$ & $0.557$ & $0.523$ & $0.132(3)$ & $0.150$ & $0.120$ \\
			\midrule
			CDP Langevin & $0.535(1)$ & $0.561$ & $0.510$ & $0.146(3)$ & $0.179$ & $0.125$ \\
			\bottomrule
			\end{tabular}
\end{table}

\begin{figure}[tb!]
 \centering
 \includegraphics[width=3.2in]{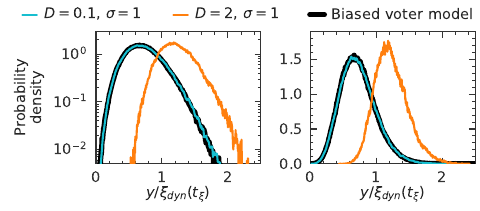}
 \caption{
  The critical height distribution for the CDP-class models with different $D/\sigma^2$ values.
  Full data are shown in Fig.~\ref{S-fig:critical_height_distributions_CDP_shift_scale}.
  }
 \label{fig:histgrams_CDP}
\end{figure}

\section{Transition probability for discrete models}

In the discrete models, $\rho((i,j),t+1)$ becomes active with the probability $p_k$, 
where $k$ is the number of connected active sites at the previous step \footnote{Precisely, $k=\sum_{i',j'\in I}\rho((i',j'),t)$, $I=\left\{(i,j),(i+\sigma,j),(i,j+\sigma),(i+\sigma,j+\sigma)\right\}$ and $\sigma=1$ ($\sigma=-1$) for the odd (even) timesteps, respectively.} (Fig.~\ref{S-fig:DK_schematic} \cite{supplementary}). 
For the bond percolation model, 
we set $p_k=1-(1-p)^k$ where $p$ is the probability with which the active states propagate \cite{henkel_2009,wang_2013}. 
For the biased voter model, we employed a parallel-update version in which the two halves of the sites are alternately updated in a checkerboard pattern \cite{supplementary}.
The site state changes to that of a randomly chosen nearest-neighbor site 
selected with biased probabilities $(1+\frac{\epsilon}{2})$ (active site) and $(1-\frac{\epsilon}{2})$ (inactive site), 
which leads to $p_k=\frac{\left(2+\epsilon\right)k}{8+2\epsilon\left(k-2\right)}$ \cite{supplementary}.

\section{Nonuniversal parameter estimation}

\subsection*{Simulation with homogeneous initial condition}
We estimated the model-dependent nonuniversal parameters for APTs by performing simulations with homogeneous initial conditions. The initial conditions were set as 
(\textit{a1}) the all-active initial condition for the bond percolation, 
(\textit{a2}) the random initial condition for the biased voter model where each site is independently set to active or inactive with probability $1/2$,
 and 
(\textit{b}) $\rhoval=\rho_\mathrm{init}=10 \text{(DP)}, 0.5 \text{(CDP)}$ for the Langevin equation. The estimated parameters are summarized in Table~\ref{S-tab:DP_nonuniversal} and \ref{S-tab:CDP_nonuniversal}. 

\subsection*{Critical parameters of DP models}
The bond percolation model showed the transition at the critical value of $p$ estimated in Ref.~\cite{wang_2013} [Table \ref{tab:critical_parameters}, Fig.~\ref{S-fig:DP_mean_rho_A_c} (a,b) \cite{supplementary}].
For the DP Langevin equation, we estimated the critical value by finding a value of $A$ with which the mean order parameter $\ave{\rhoval}$ 
decays algebraically as $\sim t^{-\betadp/\nuparadp}$ in a simulation with a homogeneous initial condition \cite{supplementary}, where $\betadp$ and $\nuparadp$ are universal exponents for the DP class [Table~\ref{tab:exponent_values}, Table \ref{tab:critical_parameters}; Fig.~\ref{S-fig:DP_mean_rho_A_c} (c-e) \cite{supplementary}]. 

\subsection*{Characteristic timescales}

For the DP models, we estimated the correlation time $\xipara$ by fitting the decay of the mean order parameter to $\propto\exp\left[-t/\xipara\right]$, and then extracted $t_0$ from the scaling form $\xipara:=t_0\epsilon^{-\nuparadp}$. The stationary density $\rhoinf$ was obtained from the long-time value of $\ave{\rhoval}$ for $\epsilon>0$ \cite{supplementary}. Rescaling $t$ and $\rhoval$ by $\xipara$ and $\rhoinf$, respectively, produced a scaling collapse of $\ave{\rhoval}$ (Fig.~\ref{S-fig:DP_rescaled_mean_rho} \cite{supplementary}).

For the CDP models, we first fitted an exponential function $C_{t0} \exp\left[{-t/\xipara}\right]$ to the tail of the mean density $1-\ave{\rhoval}$ for $\epsilon>0$, where $C_{t0}$ is a constant that depends on the value of $\epsilon$ and $\xipara$ is the correlation time. 
By plotting the values of $\xipara$ against $\epsilon$, we noticed a linear relationship between $\xipara\epsilon$ and $\log\xipara$, in agreement with the theoretical expectation \cite{janssen_2005}. We then fitted a linear function to those values for the parameters shown in Fig.~\ref{S-fig:CDP_corr_time} to estimate the parameters $C_1$ and $C_2$ for the relationship $\epsilon\xipara = C_1+C_2\ln\xipara$, which are used to calculate the correlation time for general $\epsilon$ as $\xipara = -\frac{C_2}{\epsilon}W_{-1}\left[-\frac{\epsilon}{C_2}\exp\left(-\frac{C_1}{C_2}\right)\right]$, where $W_{-1}$ is the lower branch of the Lambert $W$ function \cite{supplementary}.
The estimated parameters are summarized in Table \ref{S-tab:CDP_nonuniversal} and Fig.~\ref{S-fig:CDP_corr_time_summarized} \cite{supplementary}. Interestingly, we found that the value of $C_2$ depends on the dimensionless parameter $D/\sigma^2$ (Fig.~\ref{S-fig:CDP_corr_time_summarized} 
 \cite{supplementary}). Rescaling the mean density by the estimated correlation time collapsed the data for different $\epsilon$ values onto a single curve (Fig.~\ref{S-fig:CDP_corr_time_rescale} \cite{supplementary}).

\subsection*{Characteristic length scales}

We then estimated the correlation length $\xiperp$ as follows. 
For the DP models, we fitted an empirical function \cite{henkel_2009,takeuchi_2009} 
\begin{align}
\tilde{C}(\hat{l})=C_{s0}\hat{l}^{-(\betadp/\nuperpdp)}e^{-\hat{l}}, 
\end{align}
where $\hat{l}:=l/\xiperp$ and $\xiperp:=x_0\epsilon^{-\nuperpdp}$, 
to the stationary rescaled spatial covariance $C_s(l,t\gg\xipara)/\rhoinf^2$, where $C_s(l,t):=\Cov{\rho\left(\mathbf{x}+l\mathbf{e},t\right)}{\rhoval}$, $\Cov{A}{B}$ is the covariance between $A$ and $B$, and $\mathbf{e}$ is a unit vector (Fig.~\ref{S-fig:DP_corr_len} \cite{supplementary}). 

For the CDP models, we defined the correlation length using the dynamic correlation length at criticality ($\epsilon=0$), because the stationary state is trivial unless $\epsilon=0$.
Specifically, we fitted the scaled exact solution \cite{shimaya_2019a,krapivsky_1992} 
\begin{equation}
    C_{\mathrm{dyn}}\left[\ln(t/t'_0)\right]^{-1}\mathrm{Ei}_1\left[l^2/\xidyn{t}^2\right],
    \label{eq:CDP_xidyn}
\end{equation} where $\mathrm{Ei}_1(x):=\int_x^\infty w^{-1} e^{-w} \mathrm{d}w$, $\xidyn{t}:=\left(D't\right)^{1/\zcdp}$, and $\zcdp:=\nuparacdp/\nuperpcdp=2$
to the spatial covariance $C_s(l,t)$ at criticality to find the dynamic correlation length $\xidyn{t}$, and then defined the correlation length by $\xiperp:=\xidyn{\xipara}$ (Figs.~\ref{S-fig:CDP_corr_len_rescale} and \ref{S-fig:CDP_corr_len_summarized} \cite{supplementary}). The $C_s(l,t)$ rescaled by the scaling ansatz agreed with the exact function with the maximum deviation $\approx 0.5$ (Fig.~\ref{S-fig:CDP_corr_len_rescale} \cite{supplementary}). We found that $D'$ depends roughly linearly on $D$ as $D'\approx 7.7(1) \times D$ (Fig.~\ref{S-fig:CDP_corr_len_summarized} \cite{supplementary}).

\section{Parameter estimation for the KPZ interfaces in the long-time limit} 

We estimated the parameters for the KPZ interfaces following a standard method \cite{takeuchi_2012,takeuchi_2018,krug_1992_amplitude}. In short, we fitted a linear function to $\ave{\frac{\partial h_\xi}{\partial t_\xi}}$ versus $t_\xi^{-2/3}$ to estimate $v_\infty$ from the intercept, and similarly fitted $t_\xi^{-2/3}\cum{h_\xi}{2}$ versus $t_\xi^{-2/3}$ to estimate $\Gamma$ from the intercept, assuming that $\chi$ in Eq.~\eqref{eq:h_asymptotic} follows the GOE-TW distribution. We then calculated $A$ using $\Gamma=\frac{A^2v_\infty}{2}$ \cite{takeuchi_2018}. The estimated parameters are summarized in Fig.~\ref{S-fig:KPZ_nonuniversal} and Table \ref{S-tab:nonuniversal_params_KPZ} \cite{supplementary}. The data for $t>5\times 10^3$ were used for the fitting, and the fitting was not attempted if there were fewer than $10$ data points in the fitting range.

Specifically, for the $v_\infty$ estimation, we approximated the derivative $\ave{\frac{\partial h_\xi}{\partial t_\xi}}$ by the finite difference of the mean height $\ave{h_\xi}$ at $t_\xi$ and $t_\xi+\Delta t_\xi$ with $\Delta t_\xi< 0.38 \times t_\xi$. 
The value of $v_\infty$ was estimated by fitting a linear function to the values $\ave{\frac{\partial h_\xi}{\partial t_\xi}}$ with respect to $t_\xi^{-2/3}$ in the range $(0,w)$ and taking the value of the intercept. The value of $w$ varied from $0.15$ to $0.2$, and the overall uncertainty was derived from the combined statistical uncertainty ranges. 
We estimated the value of $\Gamma$ by fitting a linear function to $t_\xi^{-2/3}\cum{h_\xi}{2}$ versus $t_\xi^{-2/3}$
in the range $(0,w')$ and calculating $(B/\cum{\chi_1}{2})^{3/2}$, where $B$ is the intercept value and $\chi_1$ is the random variable following the Gaussian orthogonal ensemble Tracy--Widom distribution \cite{takeuchi_2012}.
The value of $w'$ was varied from $0.15$ to $0.2$ (from $0.4$ to $0.6$) in the DP (CDP) models, respectively, 
and the overall uncertainty was derived from the combined statistical uncertainty ranges. 

\subsection*{Height spatial covariance}

We observed a data collapse upon rescaling $C_{s,int}(l,t_\xi)$ and $l$ by the APT dynamical correlation length in the short-time regime, and by the KPZ-class fluctuation amplitude and correlation length in the long-time regime. In the long-time limit, the values agreed with those of the exact solution for the KPZ class with flat initial conditions, the Airy$_1$ covariance.
The Langevin equations showed the same behavior as the discrete models %
(Figs.~\ref{S-fig:off_critical_spatial_correlation_DP} and \ref{S-fig:off_critical_spatial_correlation_CDP} \cite{supplementary}).

\bibliography{citations}

\end{document}


\title{Supplementary material for \\Universal interface fluctuations in absorbing-state phase transitions}

\date{\today}

\maketitle

\onecolumngrid
\section*{Supplementary Text 1: Domany-Kinzel model and biased voter model}
\begin{figure}[!htbp]
	\centering
	\includegraphics[width=2in]{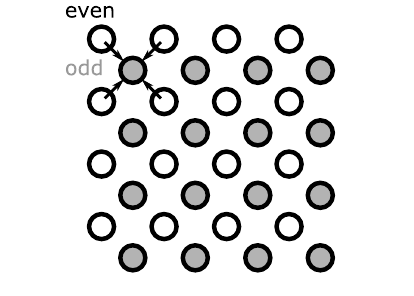}
	\caption{
		An illustration of the lattice configuration of the two-dimensional Domany-Kinzel model.
		The white and gray circles indicate the lattice sites updated at the odd and even timesteps, respectively.
		The top-left site corresponds to $(i,j)=(0,0)$ for both lattices.
		The arrows indicate the four nearest-neighbor sites that contribute to the center site (see main text).
		Periodic boundary conditions are omitted from the illustration.
	}
   \label{fig:DK_schematic}
\end{figure}

We implemented a parallel-update biased voter model as a variant of the Domany-Kinzel (DK) model as follows. 
We considered two square lattices arranged in a checkerboard pattern (white and gray sites in Fig.~\ref{fig:DK_schematic}).
In each timestep, we alternately selected the sites on one of the two lattices and updated each site's state to that of a stochastically chosen nearest-neighbor site on the other lattice.
We set the probabilities of choosing an active and an inactive site, $p_{+}$ and $p_{-}$, respectively, so that they satisfy 
$p_{+}:p_{-} = (1+\frac{\epsilon}{2}):(1-\frac{\epsilon}{2})$, thus 
\begin{align}
	p_{\pm} = \frac{2\pm\epsilon}{8+2\epsilon\left(k-2\right)},
\end{align}
where $k$ is the number of nearest-neighbor active sites.
In terms of the Domany-Kinzel model, this corresponds to choosing $p_k$ to be 
\begin{align}
	p_k=\frac{\left(2+\epsilon\right)k}{8+2\epsilon\left(k-2\right)}.
\end{align}

\section*{Supplementary Text 2: Numerical simulation of Langevin equations}

We simulated the Langevin equation
\begin{align}
	\frac{\partial \rho(\mathbf{x}, t)}{\partial t}=D \triangle\rho+A\rho-B\rho^2+\sigma\sqrt{f(\rho)}\eta(\mathbf{x}, t)
	\tag{\ref{eq:DP-CDP-Langevin}}
\end{align}
using the scheme of Ref.~\cite{dornic_2005} summarized as follows. 
First, we approximated the spatial derivative by finite differences to obtain
\begin{align}
\diff{t} \rho_{i,j}(t) = D \triangle \rho_{i,j} + A\rho_{i,j}-B\rho_{i,j}^2+\frac{\sigma}{\Delta x}\sqrt{f(\rho_{i,j})}\eta_{i,j}(t)
\label{eq:dp_langevin_descritized}
\end{align}
where 
\begin{align}
\rho_{i,j}(t)&:=\rho\left(\left(i\Delta x,j\Delta x\right),t\right),  \\
\triangle \rho_{i,j}(t) &:= \alpha_{i,j}-\frac{4 \rho_{i,j}(t)}{(\Delta x)^2}, \\
\alpha_{i,j}&:=(\Delta x)^{-2}\sum_{\substack{k=-1,1\\l=-1,1}} \rho_{i+k,j+l}(t) 
\end{align} 
and $\eta_{i,j}(t)$ is a white Gaussian noise satisfying 
$\ave{\eta_{i,j}(t)}=0$, $\ave{\eta_{i',j'}(t')\eta_{i,j}(t)}=\delta_{i',i}\delta_{j',j}\delta(t'-t)$.
The prefactor $\frac{1}{\Delta x}$ in the last term is chosen so that the stochastic term has the same distribution as the spatially averaged noise term of the continuous version, $\frac{1}{(\Delta x)^2} \int_{x}^{x+\Delta x}\rd x' \int_{y}^{y+\Delta x}\rd y' \eta(x',y',t)$.

Regarding the time discretization, it is known that a na\"{i}ve Euler-type scheme, which approximates the noise term by a Gaussian random variable, is inappropriate because it can violate the non-negativity condition $\rho(x,y,t)\ge 0$, a crucial property of the solution of the directed percolation (DP) Langevin equation.

To avoid this problem, we used the scheme proposed in Ref.~\cite{dornic_2005}, which is an operator splitting method \cite{press_2007} that splits the terms into groups, each of which can be treated exactly, and integrates the terms group by group.
The steps for a single timestep $\Delta t$ are as follows:
\begin{enumerate}
	\item In Eq.~\eqref{eq:dp_langevin_descritized}, we first integrate the terms 
	$D \triangle \rho_{i,j} + A\rho_{i,j}+\frac{\sigma}{\Delta x}\sqrt{f(\rho_{i,j})}\eta_{i,j}(t)$
	by integrating the stochastic differential equation (SDE) 
	\begin{align}
	\diff{t}  \rho(t)=\beta\rho+\tilde{\alpha}+\tilde{\sigma}\sqrt{f(\rho)}\eta_{i,j}(t)
	\label{eq:operator_splitting1}
	\end{align}
	 with the initial condition $\rho(0)=\rho_{i,j}(t)$
	 for the timestep $\Delta t$, 
	 where $\tilde{\alpha}:=D\alpha_{i,j}$ is treated as a constant,
	 $\tilde{\sigma}:=\frac{\sigma}{\Delta x}$, and 
	 $\beta:=A-\frac{4D}{(\Delta x)^2}$.
	 \begin{itemize}
		\item For the DP Langevin equation ($f(\rho)=\rho$), this step is conducted by using the exact solution for the Fokker-Plank equation of the SDE [Eq.~\eqref{eq:operator_splitting1}], which leads to the expression for $\rho^*:=\rho(\Delta t)$ \cite{dornic_2005}:
		\begin{align}	 
		 \rho^*=\mathrm{Gamma}\left[\mu+1+\mathrm{Poisson}\left[ \lambda \omega \rho(0) \right] \right]/\lambda
		\end{align}
		 where 
		 $\mathrm{Gamma}\left[z\right]$ is a random variable following the gamma distribution with the shape parameter $z$ and the scale parameter $1$,
		 $\mathrm{Poisson}\left[w\right]$  is a random variable following the Poisson distribution with the mean $w$,
		 $\mu:=-1+\frac{2\tilde{\alpha}}{\tilde{\sigma}^2}=-1+\frac{2D\alpha_{i,j}(\Delta x)^2}{\sigma^2}$, 
		 $\lambda:=\frac{2\beta}{\tilde{\sigma}^2(\omega -1)}=\frac{2\beta(\Delta x)^2}{\sigma^2(\omega -1)}$ and $\omega:=\mathrm{e}^{\beta t}$.
		 \item For the compact directed percolation (CDP) Langevin equation ($f(\rho)=\rho(1-\rho)$), we approximate the term $\sqrt{\rho(1-\rho)}$ by 
		 \begin{align}
			\begin{cases} 
				\sqrt{\rho} & (\rho<\frac{1}{2})\\ 
				\sqrt{1-\rho} & (\rho>\frac{1}{2}), 
			\end{cases}
			\label{eq:random_cdp}
		\end{align}
		following \cite{dornic_2005}, and use the same scheme as the DP case. 
		If $\rho$ exactly equals $\frac{1}{2}$ within the numerical precision, 
		either branch of Eq.~\eqref{eq:random_cdp} is chosen randomly.
	 \end{itemize}
	 \item We then integrate the remaining part $-B\rho_{i,j}^2$ for the timestep $\Delta t$ 
	 by solving $\rd_t \rho(t)=-B\rho^2$ 
	 with the initial condition $\rho(0)=\rho^*$ to obtain
	\begin{align}	 
	\rho_{i,j}(t+\Delta t)=\rho(\Delta t)=\frac{\rho^*}{1+\rho^*B\Delta t}.
	\end{align}	 
\end{enumerate}
It has been shown that this algorithm preserves the non-negativity of the solution \cite{dornic_2005}.

\subsection*{Supplementary Text 3: Details of nonuniversal parameter estimation}

\subsubsection*{Stationary density of the DP models}
To estimate $\rhoinf$ for the DP models, we defined a timescale $t_\mathrm{stat}$ substantially longer than the correlation time, and then calculated the average of $\rhoval$ for $t>t_\mathrm{stat}$, which we denote by $\ave{\rhoval}_{t>t_\mathrm{stat}}$. We varied $t_\mathrm{stat}$ from $10\times\xipara$ to $20\times\xipara$, deriving the overall uncertainty by combining the statistical uncertainty ranges ($2\times\text{standard error}$) over these intervals.

For the bond percolation model, we averaged the values of $\tilde{\rho}$ for $\epsilon\le0.01$ to estimate $\rho_\infty$. For the DP Langevin equation, we empirically fitted a linear function to the values. The estimated parameters are summarized in Table \ref{tab:DP_nonuniversal}.

\subsubsection*{Characteristic length scale of the DP models}
\begin{figure}[!htbp]
	\centering
	\includegraphics[width=3.2in]{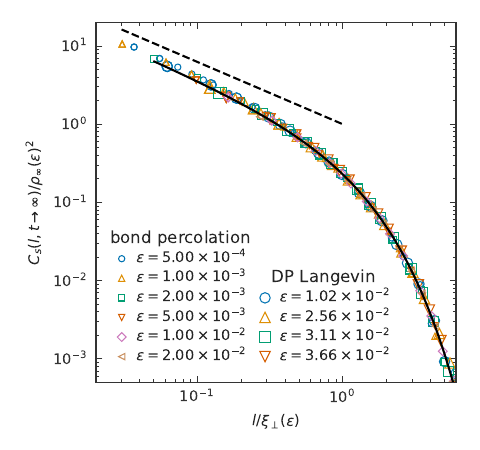}
	\caption{
		The scaling collapse plot for the spatial covariance $C_s(l,t)$ of the DP-class models 
		for the stationary timescale $t\gg\xipara$.
		The black solid line indicates the fitted spatial covariance function $\tilde{C}(l/\xiperp)$.
		The black dashed line is a guide for the eye with the exponent $l^{-\betadp/\nuperpdp}$.
	}
   \label{fig:DP_corr_len}
\end{figure}

To estimate the characteristic length $x_0$ for the DP models, we fitted the empirical function $A \tilde{l}^{-\betadp/\nuperpdp}e^{-\tilde{l}/x_0}$ to the values of $C_s(l,t=t_\mathrm{max})/\rhoinf^2$, where $\tilde{l}:=l/\epsilon^{-\nuperpdp}$, $A$ is a model-dependent constant, and $t_\mathrm{max}$ is the maximum $t$ value indicated in Table~\ref{tab:quench_params_DP}. The fitting was performed by the Levenberg-Marquardt algorithm [\texttt{scipy.optimize.curve\_fit} function in the \texttt{scipy} package (version 1.10.1)] with the initial guess $A=0.03$ and $x_0=1$ and the maximum function evaluation \texttt{maxfev=1000000}.
The universal parameter $C_{s0} = A x_0^{\beta/\nuperp}$ was estimated to be $0.61(7)$.
The estimated parameters are summarized in Table \ref{tab:DP_nonuniversal}, and the fitted function is shown in Fig.~\ref{fig:DP_corr_len}.

\newpage
\FloatBarrier
\twocolumngrid

\section{Supplemental Tables}
\begin{table}[!ht]
	\caption{\label{tab:quench_params_DP}
		(Attached as a separate file.) Parameters for the quench simulations of the DP-class models for Figs.~\ref{fig:DP_mean_rho_A_c}, \ref{fig:DP_rescaled_mean_rho}, and \ref{fig:DP_corr_len}.
		All simulations of the directed percolation were conducted with $L_x=L_y=16384$. 
		All simulations of the Langevin equation were conducted with $L_x=L_y=8192$, $D=1$, and $\sigma=1$. 
		All simulations were conducted with the double-precision float. 
		}

\end{table}

\begin{table}[!ht]
	\caption{\label{tab:quench_params_CDP}
		(Attached as a separate file.) Parameters for the quench simulations of the CDP-class models 
		for Figs.~\ref{fig:CDP_corr_time},
            \ref{fig:CDP_corr_time_summarized}, 
            \ref{fig:CDP_corr_time_rescale}.
		All simulations of the biased voter model and Langevin equation 
            were conducted with $L_x=L_y=32768$ and $L_x=L_y=8192$, respectively, with the double-precision float. 
		}
\end{table}

\begin{table}[!ht]
	\caption{\label{tab:critical_params_CDP}
		(Attached as a separate file.) The parameters for the quench simulations of the CDP-class models at criticality ($\epsilon=0$)
		for Figs.~\ref{fig:CDP_corr_len_rescale} and 
            \ref{fig:CDP_corr_len_summarized}.
		All simulations of the biased voter model and Langevin equation were conducted with $L_x=L_y=16384$ and $L_x=L_y=8192$, respectively, with the double-precision float. 
		}
\end{table}

\begin{table}[!ht]
	\caption{\label{tab:interface_params_DP}
		(Attached as a separate file) Parameters for the simulation of DP-class models with the active wall. 
		All Langevin simulations were conducted with $D=1$ and $\sigma=1$ 
		with the single-precision float. 
		}
\end{table}

\begin{table}[!ht]
	\caption{\label{tab:interface_params_CDP}
		(Attached as a separate file) Parameters for the simulation of CDP-class models with the active wall.
		All Langevin simulations were conducted with $\rho_b=1$ 
		with the single-precision float. 
		}
\end{table}

\begin{table}[!ht]
	\caption{\label{tab:DP_nonuniversal}
		Estimated nonuniversal parameters for the DP models.
            The uncertainties were derived by combining uncertainty ranges of the fitting parameters estimated for different fitting ranges, estimated values of the critical parameter, and the exponent values. 
            }
		\begin{tabular}{l|c|c|c}
			\toprule
			name & $t_0$ & $\rho_0$ & $x_0$ \\
			\midrule
			Bond percolation  & $0.463(6)$  & $1.642(5)$ & $0.21(1)$ \\
			DP Langevin       & $6.97(5)$ & $0.1684(5)$ $+0.17(1)\epsilon$ & $1.71(8)$  \\ 
			\bottomrule
		\end{tabular}
\end{table}

\begin{table}[!htbp]
	\caption{\label{tab:CDP_nonuniversal}
		Estimated nonuniversal parameters for the CDP models. 
		The values of $t'_0$ were omitted due to their large uncertainties.
            The uncertainties were derived by combining the uncertainties of the fitting parameters estimated for different fitting ranges. 
            }
		\begin{tabular}{c|c|c|c|c|c|c}
		\toprule
		model & $D$ & $\sigma$ & $C_1$ & $C_2$ & $D'$ & $C_{\mathrm{dyn}}$ \\
		\midrule
		Biased voter & -- & -- & 0.66(2) & 0.150(3) & 2.13(5) & 0.18(6) \\
		\midrule
		\multirow{8}{*}{\shortstack{CDP\\Langevin}}  
		 & 0.1 & 1 & 1.3(4) & 0.54(4) & 0.85(1)  & 0.21(4) \\
		 & 0.25 & 0.5 & 1.0(1) & 0.07(2) & 2.03(2)  & 0.10(1) \\
		 & 0.25 & 1 & 1.3(3) & 0.22(4) & 2.09(2)  & 0.16(3) \\
		 & 0.25 & 2 & 1.7(7) & 1.1(1) & 2.05(5) & 0.19(9) \\
		 & 0.5 & 1 & 1.2(1) & 0.13(2) & 4.0(3) & 0.13(4) \\
		 & 0.5 & 1.414214 & 1.2(3) & 0.29(5) & 4.0(1) & 0.15(9) \\
		 & 1 & 1 & 1.13(9) & 0.07(1) & 7.7(1) & 0.10(3) \\
		 & 2 & 1 & 1.14(7) & 0.04(1) & 14.5(5) & 0.080(4) \\
		\bottomrule
		\end{tabular}
\end{table}

 \begin{table*}[!ht]
	\caption{\label{tab:nonuniversal_params_KPZ}
		Estimated parameters of the KPZ class in the long-time regime (full table). The meaning of the column values is denoted in the caption of \ref{tab:nonuniversal_params_KPZ_full}. The values are also shown in  Fig.~\ref{fig:KPZ_nonuniversal}.
		}
		\begin{tabular}{c|c|c|c|c|c|c|c|c|c}
			\toprule
			 &  &  &  & \multicolumn{3}{c}{$v_{\infty}$} & \multicolumn{3}{c}{$\Gamma$} \\
			\cline{5-10}
			model & $D$ & $\sigma$ & $\rho_b$ & Value & Max & Min & Value & Max & Min \\
			\midrule
			Bond percolation & -- & -- & -- & $2.756(3)$ & $2.894$ & $2.627$ & $14.2(1)$ & $16.8$ & $12.1$ \\
			\midrule
			\multirow[t]{3}{*}{DP Langevin} & \multirow[t]{3}{*}{$1$} & \multirow[t]{3}{*}{$1$} 
					& $0.1$ & $2.84(1)$ & $3.00$ & $2.70$ & $15.1(5)$ & $18.5$ & $12.5$ \\
			\cline{4-10}
			&  &  & $1$ & $2.83(1)$ & $2.99$ & $2.68$ & $16(1)$ & $21$ & $12$ \\
			\cline{4-10}
			&  &  & $10$ & $2.825(3)$ & $2.968$ & $2.694$ & $15.6(4)$ & $18.8$ & $12.9$ \\
			\midrule
			Biased voter & -- & -- & -- & $0.540(3)$ & $0.556$ & $0.524$ & $0.132(3)$ & $0.149$ & $0.121$ \\
			\midrule
			\multirow[t]{8}{*}{CDP Langevin} & $0.1$ & $1$ & $1$ & $0.535(1)$ & $0.561$ & $0.510$ & $0.146(3)$ & $0.179$ & $0.125$ \\
			\cline{2-10} 
			 & \multirow[t]{3}{*}{$0.25$} & $0.5$ & $1$ & $0.5865(3)$ & $0.6207$ & $0.5543$ & $0.088(2)$ & $0.108$ & $0.068$ \\
			\cline{3-10}
			 &  & $1$ & $1$ & $0.5471(4)$ & $0.5906$ & $0.5090$ & $0.124(3)$ & $0.169$ & $0.092$ \\
			\cline{3-10}
			 &  & $2$ & $1$ & $0.560(4)$ & $0.606$ & $0.519$ & $0.149(3)$ & $0.198$ & $0.119$ \\
			\cline{2-10} 
			 & \multirow[t]{2}{*}{$0.5$} & $1$ & $1$ & $0.573(3)$ & $0.616$ & $0.537$ & $0.102(2)$ & $0.128$ & $0.083$ \\
			\cline{3-10}
			 &  & 1.414214 & $1$ & $0.579(8)$ & $0.655$ & $0.521$ & $0.128(2)$ & $0.170$ & $0.095$ \\
			\cline{2-10} 
			 & $1$ & $1$ & $1$ & $0.592(3)$ & $0.623$ & $0.557$ & $0.089(3)$ & $0.106$ & $0.070$ \\
			\cline{2-10}
			 & $2$ & $1$ & $1$ & $0.615(1)$ & $0.648$ & $0.588$ & $0.069(3)$ & $0.086$ & $0.054$ \\
			\bottomrule
			\end{tabular}
\end{table*}

\FloatBarrier
\newpage
\section{Supplemental Figures}

\begin{figure}[!htbp]
	\centering
	\includegraphics[width=3.4in]{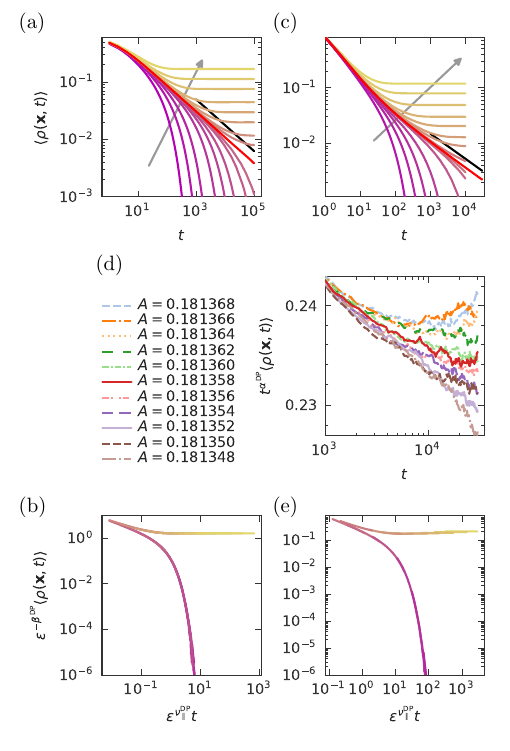}
	\caption{
		The mean value of $\rhoval$ for the quench simulations of 
		(a,b) the bond percolation model and 
		(c-e) the DP Langevin equation.
		(a,c) The original values $\ave{\rhoval}$ against time $t$. Each line corresponds to a simulation with the control parameter listed in Table \ref{tab:quench_params_DP}.
		The gray arrow indicates the direction in which the control parameter increases.
		The black line is a guide for the eye with the exponent $t^{-\alphadp}$, where $\alphadp=\betadp/\nuparadp$.
		(d) The values rescaled to compensate for the expected power law $t^{-\alphadp}$. The control parameter values are shown in the legend. All simulations used $40$ independent samples, $D=1$, $\sigma=1$, and $L_x=L_y=8192$. The critical point and its uncertainty were estimated by identifying the parameter range over which the values do not significantly increase or decrease.
		(b,e) Scaling collapse using the estimated transition-point values. The colors are the same as those used in (a,d).
	}
   \label{fig:DP_mean_rho_A_c}
\end{figure}

\begin{figure}[!htbp]
	\centering
	\includegraphics[width=3.2in]{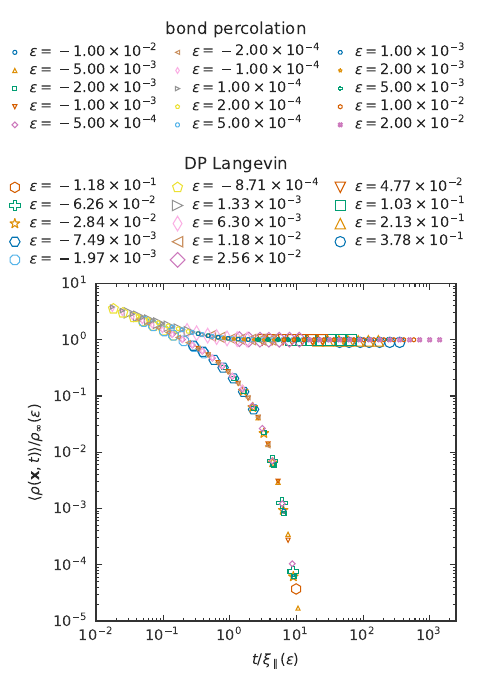}
	\caption{
		The scaling collapse plot for the mean $\rhoval$ for the quench simulations of the DP models.
	}
   \label{fig:DP_rescaled_mean_rho}
\end{figure}

\begin{figure}[!htbp]
	\centering
	\includegraphics[width=3.2in]{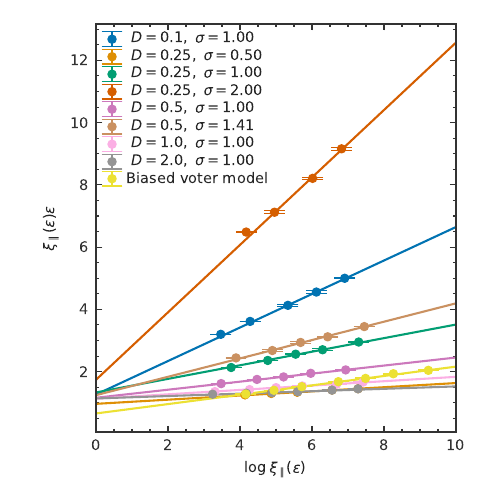}
	\caption{
		Estimation of the correlation time $\xipara$ for the CDP-class models. The lines are linear fits.
		The error bars indicate the overall uncertainty in the estimation of $\xipara$.
	}
   \label{fig:CDP_corr_time}
\end{figure}

\begin{figure}[!htbp]
	\centering
	\includegraphics[width=3.2in]{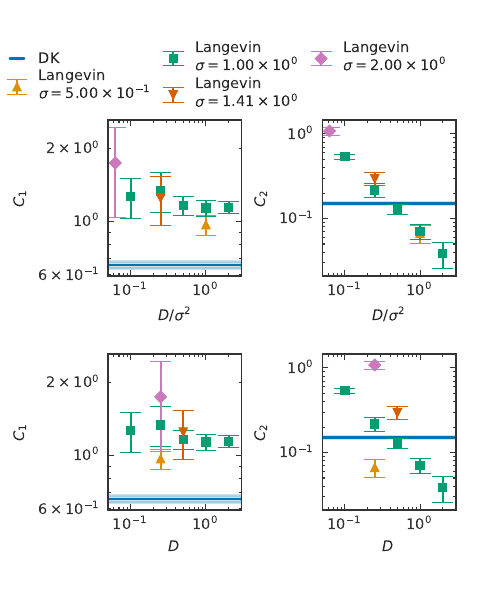}
	\caption{ 
		The estimated parameters for the CDP correlation time plotted against $D/\sigma^2$ and $D$.
		The points and error bars indicate the estimated values and uncertainties for the CDP Langevin equation, respectively.
		The blue horizontal solid line and shaded area indicate the value and uncertainty estimated for the DK model, respectively. 
	}
   \label{fig:CDP_corr_time_summarized}
\end{figure}

\begin{figure*}[!htbp]
	\includegraphics[width=6.4in]{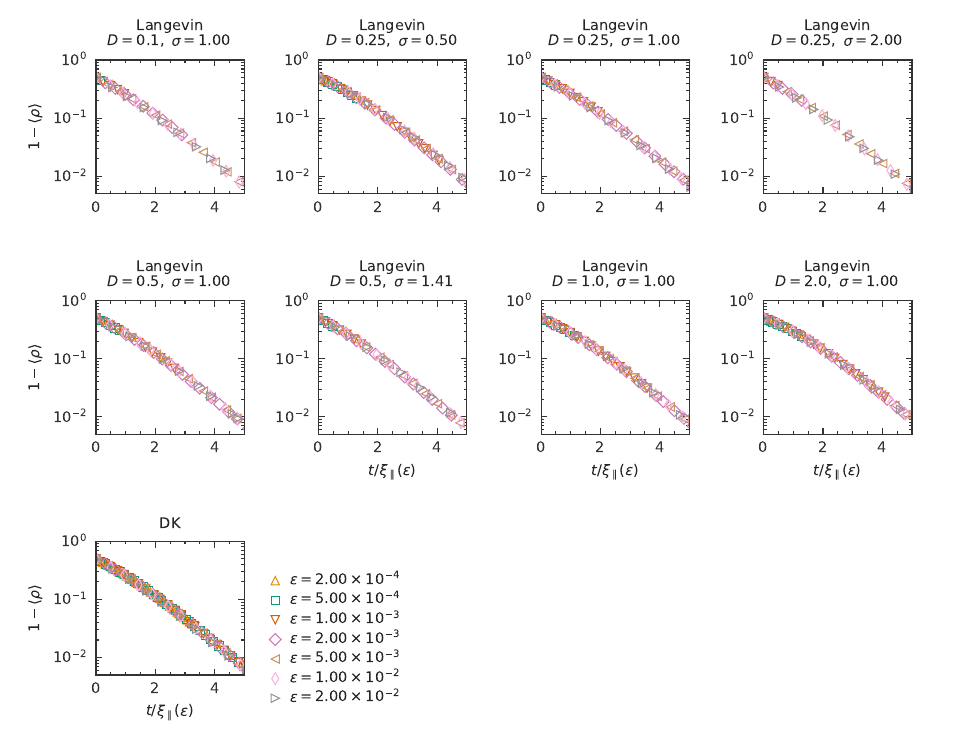}
	\caption{
		The scaling collapse plot for the mean $\rhoval$ for the quench simulations of CDP-class models.
	}
   \label{fig:CDP_corr_time_rescale}
\end{figure*}

\begin{figure}[!htbp]
	\centering
	\includegraphics[width=3.2in]{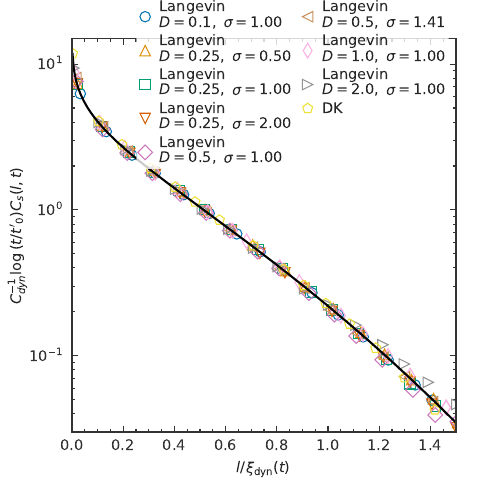}
	\caption{
		The scaling collapse plot for the spatial covariance $C_s(l,t)$ of the CDP-class models at criticality $\epsilon=0$.
		The black solid lines indicate the exact function $\mathrm{Ei}_1\left[\left(l/\xidyn{\epsilon}\right)^2\right]$. 
	}
   \label{fig:CDP_corr_len_rescale}
\end{figure}

\begin{figure}[!htbp]
	\centering
	\includegraphics[width=3.2in]{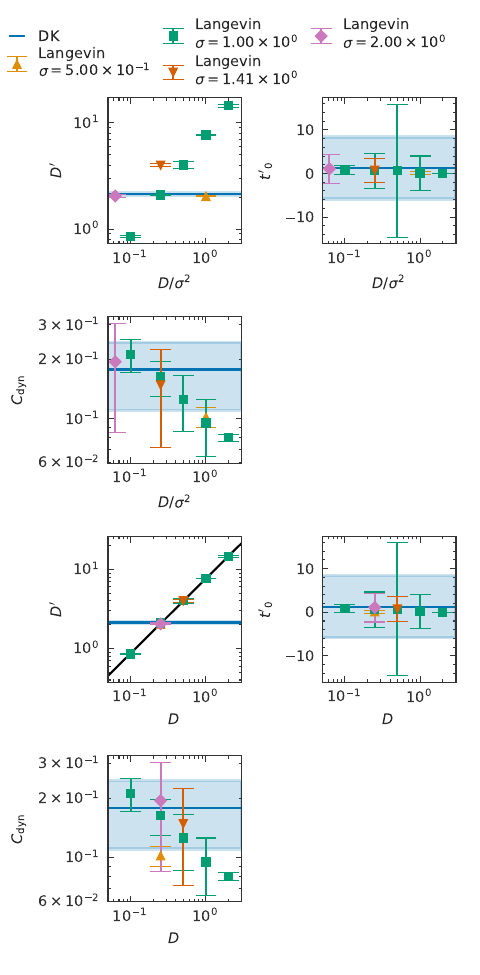}
	\caption{ 
		The estimated parameters for the dynamical correlation length for the CDP-class models, plotted against $D/b$ and $D$.
		The blue horizontal solid line and shaded area indicate the value and uncertainty estimated for the DK model, respectively. 
		The black line in the plot of $D'$ against $D$ is the result of the linear fit on a log scale, $D' = 7.7 (1) \times D^{0.95(2)}$.  
	}
   \label{fig:CDP_corr_len_summarized}
\end{figure}

\begin{figure*}[!htbp]
 \centering
 \includegraphics[width=7in]{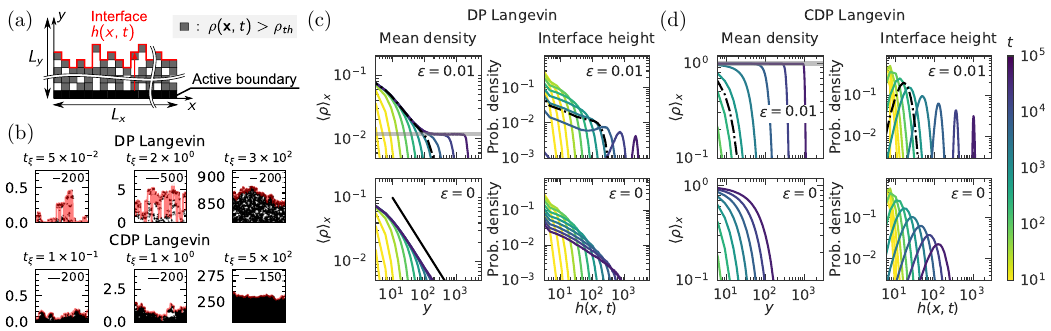}
 \caption{\label{fig:intro-langevin}
 Schematic and qualitative illustration of the interface growth process in the Langevin equation.
 (a) An illustration of the height definition with the active boundary.
 (b) Example snapshots of the active sites and the interface (red line) for the DP (upper) and CDP (lower) Langevin equations.
 The axis label and scale bar indicate the length rescaled by the correlation length $\xiperp$ and the actual length, respectively. The values of $\epsilon$ and the rescaled time $t_\xi=t/\xipara$ are indicated in each plot.
 (c,d) The spanwise-averaged density $\spanwiserho$ (left) and the height probability distribution (right) for the DP (c) and CDP (d) Langevin equations.
 The top and bottom plots are the off-critical and critical cases, respectively. %
 The dash-dot lines show the data at the correlation time $t\approx\xipara$.
 The gray lines in the off-critical mean-density plots indicate $\rhoinf$ (DP) and $1$ (CDP). The black line in the inset of the left plot is a guide for the eye for the power law $\spanwiserho\propto{}y^{-\betadp/\nuperp}$.
 For (b--d), parameters are set to $D=\sigma=1$, $\rho_b=0.1$, and $\rhoth=0.1$ for the DP model, and $D=0.1$, $\sigma=1.0$, $\rho_b=1$, and $\rhoth=0$ for the CDP model. The other parameters are summarized in Tables~\ref{tab:interface_params_DP} and \ref{tab:interface_params_CDP}.}
\end{figure*}

\begin{figure}[!htbp]
	\centering
 	\includegraphics[width=3.2in]{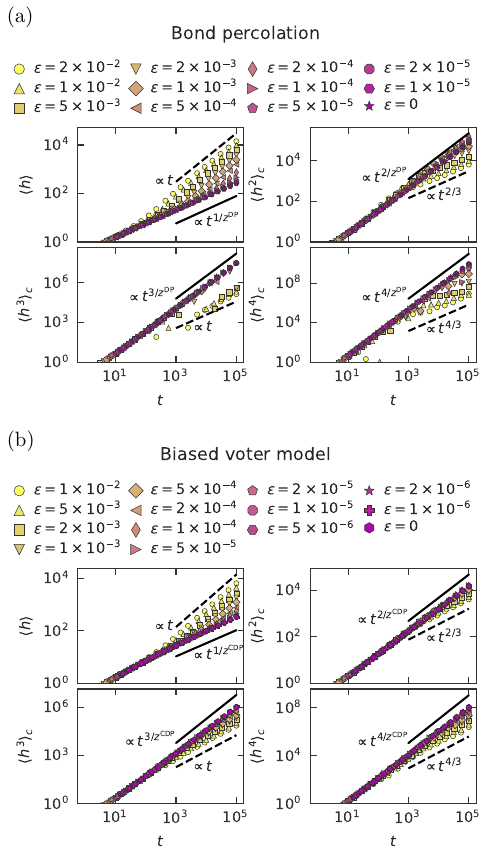}
	\caption{Cumulants of the height distribution $\cum{h_\xi}{k}$ ($k=1,2,3,4$) plotted against the raw time $t$ for the (a) bond percolation and (b) biased voter models. The solid and dashed lines are guides for the eye for the APT and KPZ exponents, respectively (see the main text).
	}
	\label{fig:raw_cumulants_DK}
\end{figure}

\begin{figure*}[!htbp]
	\centering
 	\includegraphics[width=3.2in]{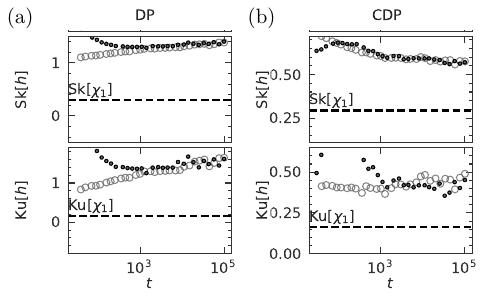}
	\caption{Cumulant ratios of the height distribution for the (a) DP- and (b) CDP-class models. The plots show the skewness $\Sk{h}$ and the kurtosis $\Ku{h}$ for the critical cases. The larger unfilled and smaller filled symbols are for the discrete models and the Langevin equations ($\rho_b=0.1$ and $\rhoth=0.1$ for the DP Langevin equation, and $\rhoth=0$, $D=0.1$, $\sigma=1.0$ for the CDP Langevin equation), respectively.
	The dashed lines show the theoretical values for the GOE-TW distribution, $\chi_1$.}
	\label{fig:critical_skew_kurtosis}
\end{figure*}

\begin{figure*}[!htbp]
	\centering
 	\includegraphics[width=6.4in]{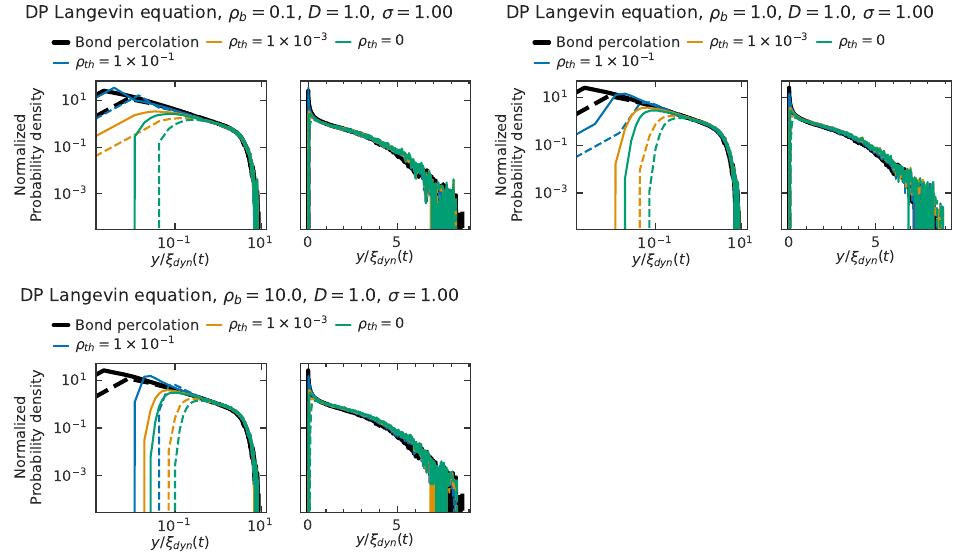}
	\caption{The height distributions for the DP-class models at criticality ($\epsilon=0$), 
	plotted with different $\rho_b$, $\rhoth$ and $t$.
	The dashed and solid lines show the data at $t=1\times10^4$ and $t=1\times10^5$, respectively.}
	\label{fig:critical_height_distributions_DP}
\end{figure*}

\begin{figure*}[!htbp]
	\centering
 	\includegraphics[width=6.4in]{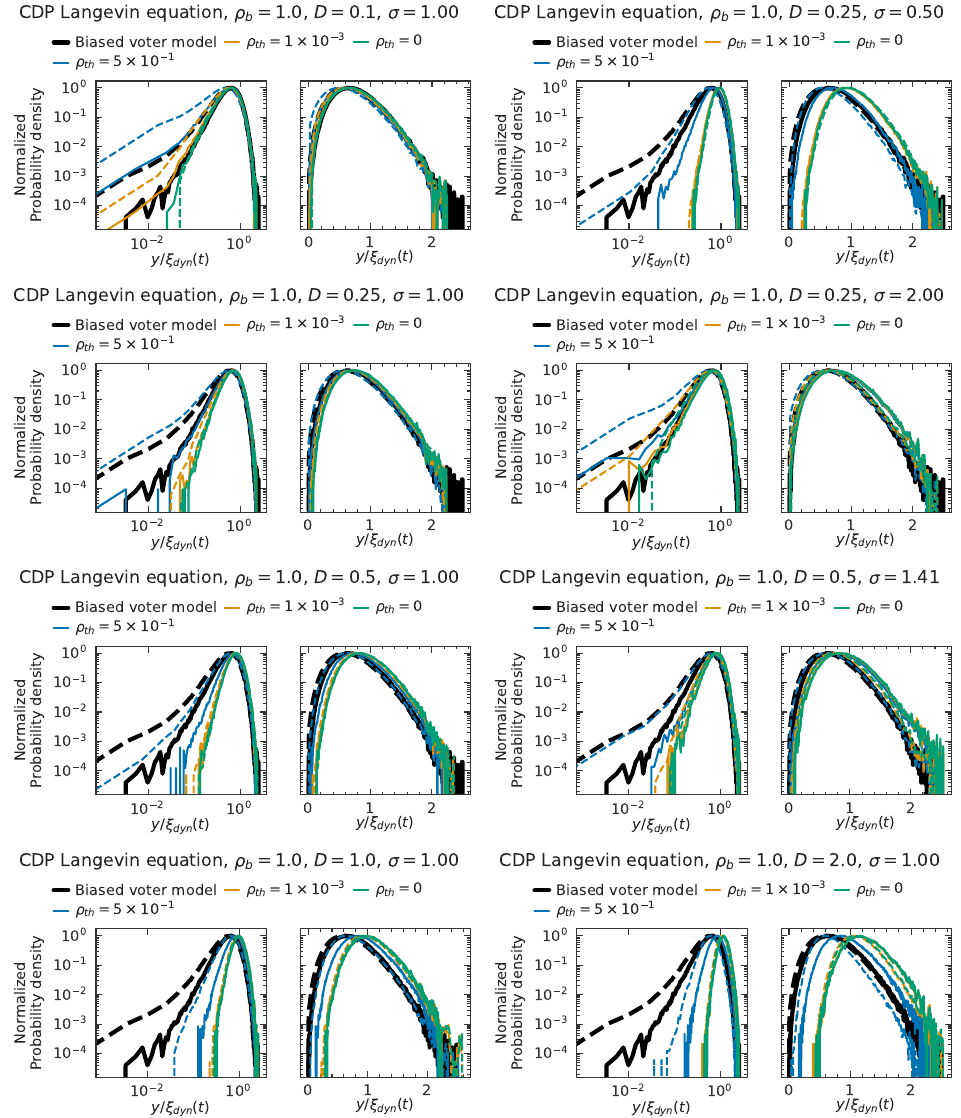}
	\caption{The height distributions for the CDP-class models at criticality ($\epsilon=0$), 
	plotted with different $D$, $\sigma$, $\rhoth$ and $t$.
	The dashed and solid lines show the data at $t=1\times10^4$ and $t=1\times10^5$, respectively.}
	\label{fig:critical_height_distributions_CDP}
\end{figure*}

\begin{figure}[!htbp]
	\centering
 	\includegraphics[width=3.2in]{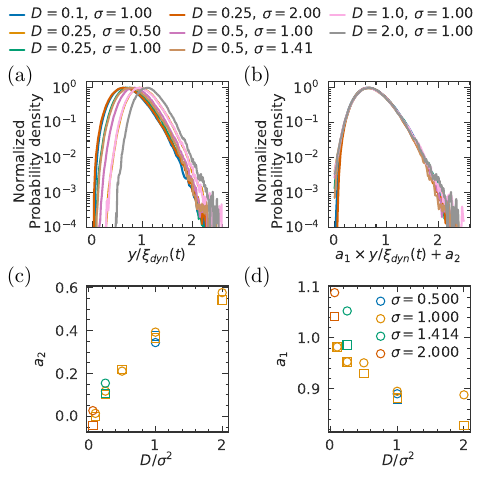}
	\caption{
		(a) Height distributions for the CDP-class models at criticality ($\epsilon=0$) with $\rhoth=0$, normalized to unity at the peak.
		(b) The same data as in (a), shifted by $a_1$ and scaled by $a_2$, where $a_1$ and $a_2$ were estimated by minimizing the squared error with respect to the histogram of the DK model using the Nelder-Mead method via the \texttt{lmfit} package \cite{newville_2024}.
		(c,d) The estimated values of $a_1$ and $a_2$ plotted against $D/\sigma^2$.
		Squares and circles correspond to $t=1\times10^4$ and $t=1\times10^5$, respectively.
	}
	\label{fig:critical_height_distributions_CDP_shift_scale}
\end{figure}

\begin{figure}[!htbp]
	\centering
 	\includegraphics[width=3.2in]{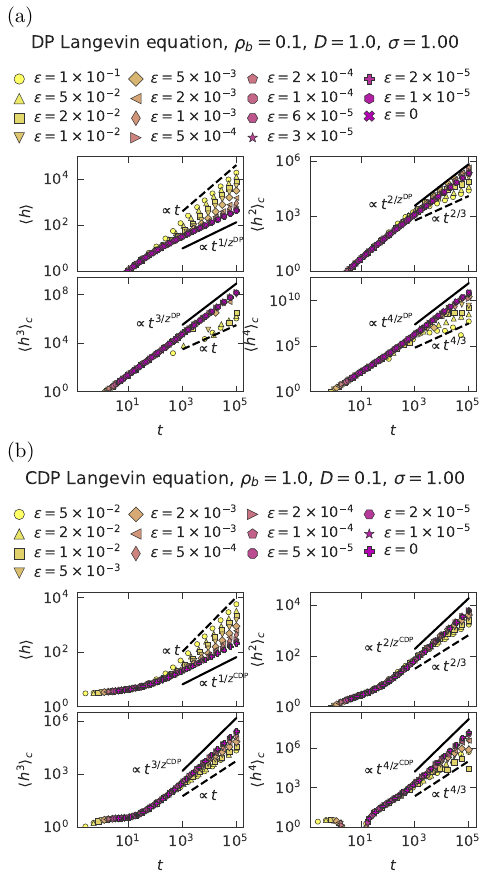}
	\caption{Cumulants of the height distribution $\cum{h_\xi}{k}$ ($k=1,2,3,4$) plotted against the raw time $t$ for the (a) DP and (b) CDP Langevin equations. The solid and dashed lines are guides for the eye for the APT and KPZ exponents, respectively (see the main text). Parameters are set to $D=\sigma=1$, $\rho_b=0.1$, and $\rhoth=0.1$ for the DP model, and $D=0.1$, $\sigma=1.0$, $\rho_b=1$, and $\rhoth=0$ for the CDP model. The other parameters are summarized in Tables~\ref{tab:interface_params_DP} and \ref{tab:interface_params_CDP}.\label{fig:raw_cumulants_langevin}}
\end{figure}

\begin{figure}[!htbp]
	\centering
 	\includegraphics[width=3.2in]{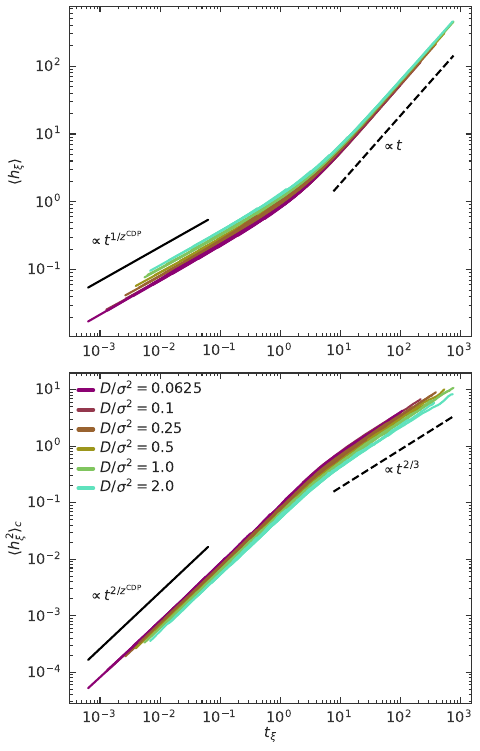}
	\caption{First and second cumulants of the height distribution for the CDP-class models. Data with $\rho_b=1.0$ and $\epsilon \le 0.01$ are plotted. The solid and dashed lines are guides for the eye for the APT and KPZ exponents (see the main text), respectively.}
	\label{fig:CDP_cumulants_overlap}
\end{figure}

\begin{figure*}[!htbp]
	\centering
	\includegraphics[width=6.4in]{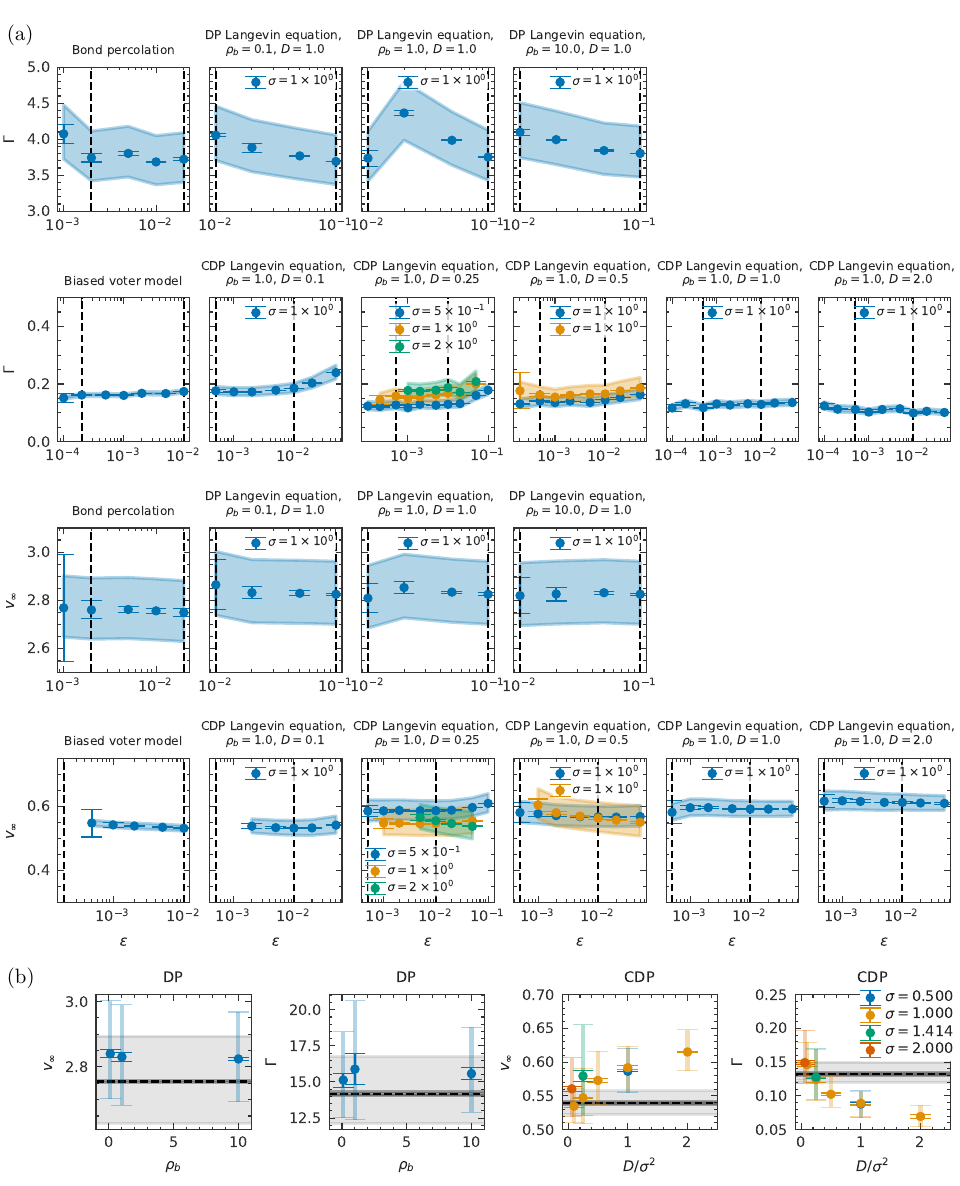}
		\caption{Estimated parameters for the KPZ interfaces, $v_\infty$ and $\Gamma$.
	Panel (a) shows the values for different $\epsilon$,
	and (b) shows the estimated values for $\epsilon\to 0$.
	The stochastic uncertainty and the systematic uncertainty due to the uncertainty in the nonuniversal parameter estimation are shown by the error bars and the shaded area in (a),
	and by the error bars with dark and light colors in (b), respectively.
	The $\epsilon$ ranges used to estimate the values in (b) are indicated by two dashed lines in (a).
	In (b), the values, stochastic uncertainty, and systematic uncertainty for the DK models are shown by the dashed line, the dark gray shaded area, and the light gray shaded area, respectively.
	The parameters used in the simulations are summarized in Tables~\ref{tab:interface_params_DP} and \ref{tab:interface_params_CDP}.
	}
\label{fig:KPZ_nonuniversal}
\end{figure*}

\begin{figure*}[!htbp]
	\centering
 	\includegraphics[width=6.4in]{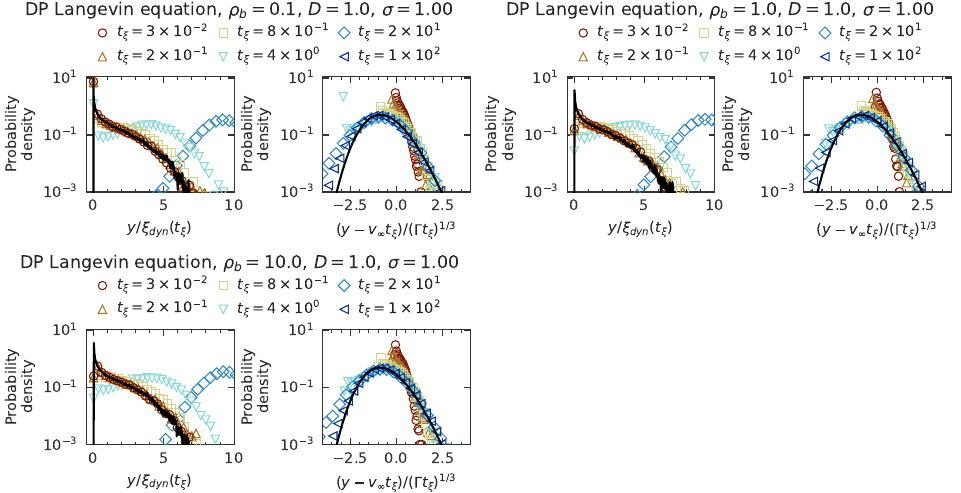}
	\caption{Height distributions for the DP models, plotted at different $t_\xi$. The black solid lines indicate the data for the critical case (left plots) and the exact solution for flat KPZ interfaces (right plots) \cite{Prahofer.Spohn-Table}, respectively.
	The parameters used in the simulations are summarized in Tables~\ref{tab:interface_params_DP} and \ref{tab:interface_params_CDP}.}
	\label{fig:off_critical_height_distributions_DP}
\end{figure*}

\begin{figure*}[!htbp]
	\centering
 	\includegraphics[width=6.4in]{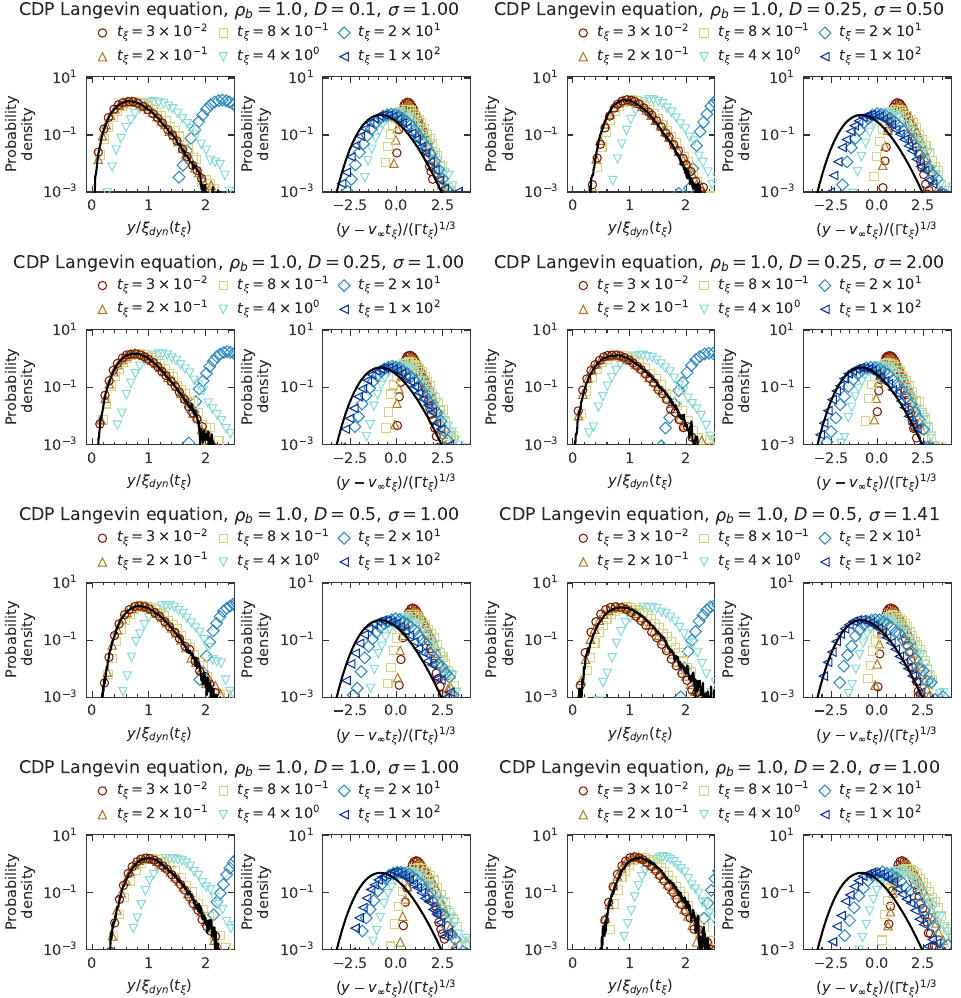}
	\caption{Height distributions for the CDP models, plotted at different $t_\xi$.
 The black solid lines indicate the data for the critical case (left plots) and the exact solution for flat KPZ interfaces (right plots) \cite{Prahofer.Spohn-Table}, respectively.
	The parameters used in the simulations are summarized in Tables~\ref{tab:interface_params_DP} and \ref{tab:interface_params_CDP}.}
	\label{fig:off_critical_height_distributions_CDP}
\end{figure*}

\begin{figure*}[!htbp]
	\centering
 	\includegraphics[width=6.4in]{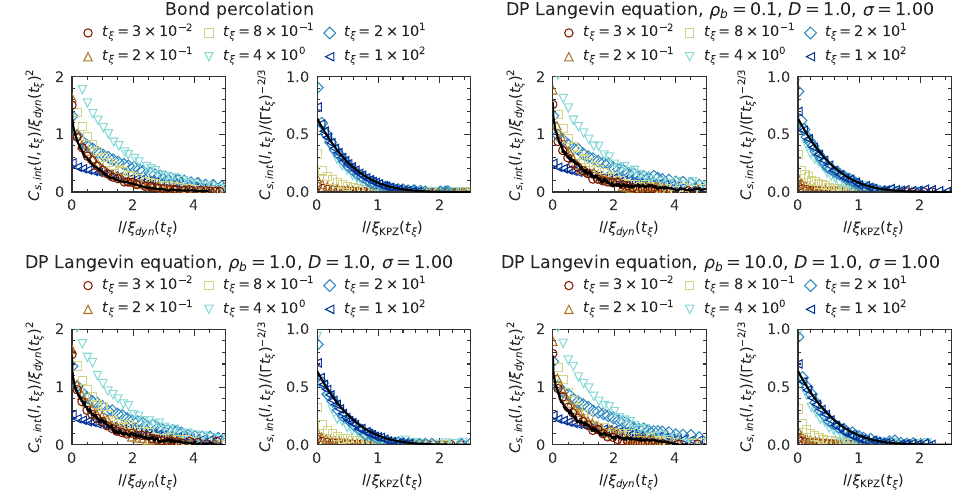}
	\caption{Height spatial covariance for the DP models, plotted at different $t_\xi$.
 The black solid lines indicate the data for the critical case (left plots) and the exact solution for flat KPZ interfaces (right plots) \cite{Bornemann-MC2010}, respectively.
 }
	\label{fig:off_critical_spatial_correlation_DP}
\end{figure*}

\begin{figure*}[!htbp]
	\centering
 	\includegraphics[width=6.4in]{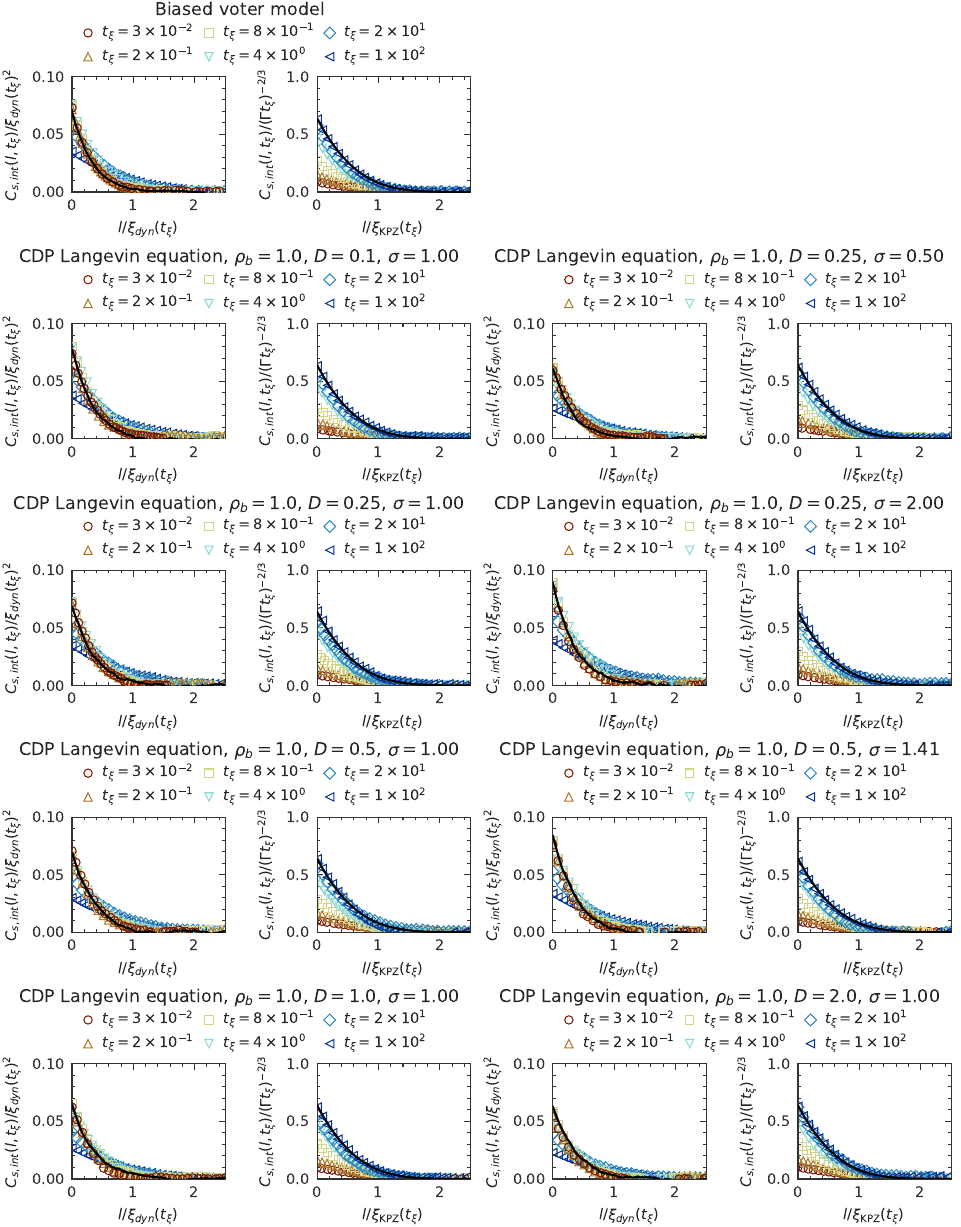}
	\caption{Height spatial covariance for the CDP models, plotted at different $t_\xi$.
 The black solid lines indicate the data for the critical case (left plots) and the exact solution for flat KPZ interfaces (right plots) \cite{Bornemann-MC2010}, respectively.}
	\label{fig:off_critical_spatial_correlation_CDP}
\end{figure*}

\FloatBarrier
\bibliography{citations}